\documentclass[reprint,prd,nofootinbib,aps]{revtex4-2}
\usepackage{amsmath,amssymb,amsfonts,subfig,graphicx,nicefrac}
\usepackage[T1]{fontenc}

\newcommand{\re}[1]{(\ref{eq:#1})}

\let\vrho\rho
\def\phi{\varphi}

\renewcommand{\rho}{\varrho}
\def\d{\mathrm{d}}
\def\p{\partial}

\def\sign{\mathop{\rm sign}}

\renewcommand{\vec}[1]{\boldsymbol{#1}}

\def\={\discretionary{-}{-}{-}}

{\vskip 3pt plus 1pt minus 1pt\begin{normalsize}}%
{\par\end{normalsize}\vskip 3pt plus 1pt minus 1pt}

\newcommand{\rz}{\rho_0}
\newcommand{\zz}{z_0}
\newcommand{\ka}{a}
\newcommand{\Krho}{\vrho}
\newcommand{\Krhocc}{\bar{\vrho}}
\newcommand{\overbar}[1]{\mkern 1.5mu\overline{\mkern-1.5mu#1\mkern-1.5mu}\mkern 1.5mu}
\newcommand{\debcc}{\bar{\psi}}

\newcommand{\thorn}{\text{\th}}

\begin{document}
\title{Debye superpotential for charged rings or circular currents on Kerr black hole}
\author{David Kofro\v{n}}\email{d.kofron@gmail.com}
\author{Petr Kotla\v{r}\'ik}\email{kotlarik.petr@gmail.com}
\affiliation{
Institute of Theoretical Physics, Faculty of Mathematics and Physics,\\
Charles University,\\
V Hole\v{s}ovi\v{c}k\'{a}ch 2, 180\,00 Prague 8, Czech Republic}


\begin{abstract}
    We provide an explicit, closed and compact expression for the Debye superpotential of a circular source. This superpotential is obtained by integrating the Green function of Teukolsky Master Equation (TME). The Debye potential itself is then, for a particular configuration, calculated in the same manner as the $\phi_0$ field component is calculated from the Green function of the TME --- by convolution of the Green function with sources.
    This way we provide an exact field of charged ring and circular current on the Kerr background, finalizing thus the work of Linet.
\end{abstract}
\maketitle

\section{Introduction}
The test electromagnetic fields on a rotating black hole -- a Kerr black hole \cite{Kerr1963} -- background are of perpetual interest for their astrophysical importance; for an overview see \cite{Punsly2009}. Fields of stationary and axisymmetric charge/current configurations attract our attention for the fact that they can represent (simplified) models of electromagnetic fields generated by accretion discs.

Yet, the task to solve Maxwell's equations on a Kerr background is highly nontrivial. 

The most fruitful approach is a special tetrad formulation based on null tetrad --- Newman\,--\,Penrose (NP) formalism \cite{Newman1962} and its refinement Geroch\,--\,Held\,--\,Penrose (GHP) formalism \cite{Geroch1973}. Then the Maxwell field equations (ME) are four coupled $1^\text{st}$ order PDEs for complex scalars $\phi_0,\,\phi_1$ and $\phi_2$.

Due to the special algebraic properties of type D spacetimes --- of which the Kerr solution is a prominent member ---  the Maxwell equations can be decoupled and cast in three $2^\textrm{nd}$ order partial differential equations for respective NP field components. 
Equations for $\phi_0$ and $\phi_2$ (so-called TMEs) were found in 1972 by Teukolsky \cite{Teukolsky1972,Teukolsky1973} while the equation for $\phi_1$ (so-called FIE) was found by Fackerell and Ipser in 1971 \cite{Fackerell1972} and elaborated recently in \cite{Jezierski2016}.
In fact, the TMEs are more general since they govern the behaviour of a test field of arbitrary spin and has been extensively studied. 

The TMEs allow us to seek for solution by method of separation of variables and therefore are widely used. Whether or not are the NP scalars components of the same field (we have decoupled equations!) is answered by the Teukolsky\,--\,Starobinsky identities (TSI) \cite{Starobinski1973,Press1973,Teukolsky1974}.

The task of finding an electromagnetic field of charged ring or circular current has been pursued by many relativist during 60's and 70's in progressively more general setting \cite{Petterson1974,Chitre1975,Petterson1975,Linet1976,Bicak1977,Bicak1976}. The very first attempts started with a classical 4-potential formulation but soon the NP approach attracted more attention. TMEs are separable --- thus it is easy to find a solution of $\phi_0,\phi_2$ corresponding to a given source in a form of infinite series. Then the remaining NP component $\phi_1$ has to be solved from ME directly.

Yet another general approach for solving test fields of arbitrary spin on type D backgrounds is to introduce the Debye potentials. A single complex scalar function (the only independent component of Hertz potential in a particular gauge) is enough to describe the whole test field. This approach has been introduced in the realms of general relativity by Cohen and Kegeles \cite{Cohen1974a,Cohen1974b}, later elaborated in \cite{Wald1978,Kegeles1979} and recently developed and explained in terms of fundamental spinor operators by Acksteiner, Andersson and B\"ackdahl in \cite{Andersson2015,Aksteiner2019}.

The Debye potentials were used by Linet in 1979 \cite{Linet1979} for construction of the electromagnetic field of a stationary axisymmetric field on Kerr background --- the theory was established 43 years ago, but no explicit results were given. One is not surprised because already the simplest possible textbook example --- a current loop in flat spacetime --- is nontrivial since it contains elliptic integrals. This is where we are going to proceed further.

The paper is organized as follows. We briefly introduce the Kerr metric and the Kinnersley tetrad in Sec. \ref{sec:Kerr} to set up the background. The spin coefficients are for the sake of brevity listed in Appendix \ref{app:Kerr}. For the same reason the congruence of zero angular momentum (ZAMO) observers, which we will later use for splitting the electromagnetic field into the electric $\vec{E}=\vec{u}\cdot\vec{F}$ and magnetic $\vec{B}=\vec{u}\cdot\star\vec{F}$ field, is introduced in Appendix \ref{app:ZAMO}. And the elliptic integrals are defined in Appendix \ref{app:ellitptic}.

We shortly introduce the TMEs in Sec. \ref{sec:TME} and the Debye potentials in Sec. \ref{sec:Debye}. A very short introduction of NP and GHP formalism can be found in Appendix \ref{app:NPGHP}. These  standard methods are in details covered in \cite{Newman1962,Geroch1973,Penrose1984,Andersson2015}. 

In Sec. \ref{sec:GASP} we shortly recall the results derived by Linet \cite{Linet1977,Linet1979}. He cast the TMEs under the assumptions of stationarity and axisymmetry into the form of generalized Laplace equations and provided Green functions. He has also shown, how to obtain the Debye potential for such fields -- using the generalized axially symmetric potential (GASP) theory. It is easy to obtain the values of potential (which is a solution of the Laplace equation) on the axis; then, the solution on the whole space is defined as a particular integral.

In Sec. \ref{sec:Int} we present the analytic solution of the Debye potential (which we call the superpotential). This superpotential gives rise to to a field with $\phi_0$ given by the Green function of the Teukolsky operator. We discuss the structure of discontinuities which we found in this superpotential and their significance. The importance of this result is clear: a closed compact analytical formula seems to be much better than an infinite series expansion (which is difficult to treat numerically close to the radius at which the source is located). 

In Sec. \ref{sec:superpotential} we discuss the properties and charge induced by this superpotential on the Kerr black hole. Sections \ref{sec:ring} and \ref{sec:loop} are devoted to presentation of realistic physical fields of given sources: a charged ring or a current loop. We numerically check our results against the series expansion solutions presented in \cite{Bicak1976}. Again, for the sake of compactness the reader is asked to refer to this paper for particular coefficients of the series.

\section{Kerr black hole}
\label{sec:Kerr}
One of the most fundamental solution of the vacuum Einstein field equations --- the rotating black hole --- was discovered in 1963 by Roy Kerr \cite{Kerr1963}. Recent historical reviews can be found in \cite{Wiltshire2009,Teukolsky2015}.

We adopt the signature convention $(-,+,+,+)$ and then metric itself in Boyer\,--\,Lindquist coordinates reads
\begin{multline}
\vec{\d} s^2 = -\frac{\Delta}{\Sigma} \left( \vec{\d} t -\ka\sin^2\theta\,\vec{\d}\phi \right)^2 +\frac{\Sigma}{\Delta}\,\vec{\d} r^2 \\
+ \Sigma \, \vec{\d}\theta^2  
+\frac{\sin^2\theta}{\Sigma}\Bigl( \left( \ka^2+r^2 \right)\vec{\d}\phi - \ka\,\vec{\d} t \Bigr)^2\,,
\label{eq:KerrMetric}
\end{multline}
with the standard definitions 
\begin{align}
\Delta&=r^2-2Mr+\ka^2=(r-r_p)(r-r_m)\,,\\
\Sigma&=\vrho\bar{\vrho}=r^2+\ka^2\cos^2\theta\,\\
\vrho&=r-i\ka\cos\theta\,.
\end{align} 
The parameters have the following meaning: $M$ is the mass of the black hole, $Ma$ is its angular momentum, $r_p$ is the position of outer black hole horizon whereas $r_m$ is the position of inner black hole horizon. We will also frequently use parameter $\beta$ which we define as\footnote{Out of parameters $r_p,r_m,M,a,\beta$ only two are independent.}
\begin{equation}
\beta=\sqrt{M^2-\ka^2}=(r_p-r_m)/2\,.
\end{equation}

The Kinnersley NP\footnote{Notice the boost given by $\sqrt{2}$ in contrast to standard textbook form. This makes the resulting expressions in terms of the Debye potentials to appear `more symmetrical'.} tetrad $(\vec{l},\,\vec{m},\,\bar{\vec{m}},\,\vec{n})$ adapted to the principal null directions of the Weyl tensor reads as follows
\begin{equation}
\begin{alignedat}{1}
\vec{l} &= \frac{1}{\sqrt{2}\,\Delta}\left[ \left( r^2+\ka^2 \right)\vec{\p_t}+\Delta\,\vec{\p_r} + \ka\,\vec{\p_\phi} \right]\,, \\
\vec{n} &= \frac{1}{\sqrt{2}\,\Sigma}\left[ \left( r^2+\ka^2 \right)\vec{\p_t}-\Delta\,\vec{\p_r} + \ka\,\vec{\p_\phi} \right]\,, \\
\vec{m} &= \frac{1}{\sqrt{2}\,\bar{\vrho}}\bigl( i\ka\sin\theta\,\vec{\p_t}+\vec{\p_\theta} +i\csc\theta\,\vec{\p_\phi} \bigr)\,.
\end{alignedat}
\label{eq:NPtetrad}
\end{equation}

Total electric charge $Q_e$ and magnetic charge $Q_m$ can be calculated by integrating two-form\footnote{Hodge dual of a two form is defined as
$$(\star \vec{F})_{ab}=\frac{1}{2}\vec{\epsilon}_{ab}^{\phantom{ab}cd}\vec{F}_{cd}\,,$$
where $\vec{\epsilon}$ is a volume element.} $\vec{F}^*=\vec{F}-i\star\vec{F}$ over a closed 2-surface. This yields
\begin{equation}
iQ_e-Q_m = \frac{1}{4\pi} \oint \vec{F}^*\,.
\label{eq:}
\end{equation}
After standard reconstruction of $\vec{F}^*$ from the NP components and the NP tetrad we get for surfaces of constant $t$ and $r$  following form of the Gauss law:
\begin{multline}
iQ_e-Q_m =  \frac{1}{2}\int_0^\pi -\Krho\,\ka\sin^2\theta \,\phi_2 \\
-2i\sin\theta \left( r^2+\ka^2 \right) \phi_1  
+\frac{\ka\Delta\sin^2\theta}{\Krho}\,\phi_0
\ \d\theta \,,
\label{eq:totQ}
\end{multline}
where we already anticipated axial symmetry.

We will also employ the Weyl coordinates which  are introduced as\footnote{Notice, that \emph{prime} is either GHP operation (when connected with spin coefficients or directional derivative operator) or standard notation for integrating parameters or it denotes  differentiation with respect to $r$ coordinate. We believe that its meaning is clear from the context.}
\begin{align}
    z&=\nicefrac{1}{2} \Delta'(r)\cos\theta\,, &
    \rho&=\sqrt{\Delta}\,\sin\theta \,.
\end{align}

\section{Teukolsky Master Equation}
\label{sec:TME}

Let us write down TME \cite{Teukolsky1972} for $\phi_0$ in the GHP formalism as
\begin{equation}
\left[\left(\thorn-\bar{\rho}-2\rho\right)\left(\thorn'-\rho'\right)-\left(\eth-\bar{\tau}'-2\tau\right)\left(\eth'-\tau'\right)\right]\phi_0 = J_0\,,
\label{eq:TME}
\end{equation}
where the sources are encoded in $J_0$ which is given in terms of projections of the four-current onto the null tetrad as
\begin{align}
    J_0&=\left(\eth-2\tau-\bar{\tau}'\right)J_l 
    - \left(\thorn -2\rho-\bar{\rho}\right)J_m\,.
\end{align}

Once the Green function $G$ is known the field of particular sources is then given by convolution of this Green function $G$ with the particular source terms $J_0$ \cite{Linet1977} as
\begin{multline}
\phi_0 = \int_0^\infty \int_0^\pi G(r,\theta,r',\theta')J_0(r',\theta',r_0,\theta_0) \times \\
\Sigma(r',\theta')\sin\theta'\,\d r'\d\theta' \,.
\label{eq:convolution}
\end{multline}

The Green function of the Teukolsky operator (the one on the l.h.s. of Eq. (\ref{eq:TME})) is easy to integrate and will be provided explicitly in the next Section. 

To know the whole electromagnetic field, one has to seek for $\phi_1$ as well. And this task is considerably more difficult. We can either 
(a) directly solve the ME in NP formalism --- which are presented in Appendix \ref{app:ME} in a simplified version for stationary and axially symmetric field; or 
(b) use the Debye potentials for the electromagnetic field. We will pursue the latter approach in Section \ref{sec:Int}.

\section{Debye potential}
\label{sec:Debye}

There exist three distinct possibilities how to choose Debye potential for the electromagnetic field. We adhere to the most common one --- a complex GHP scalar function $\debcc$ of GHP weight $[0,-2]$ which solves the the Debye equation. This equation in GHP formalism can be written as
\begin{align}
\left[\left(\thorn'-\rho'\right)\left(\thorn+\bar{\rho}\right)-\left(\eth-\tau\right)\left(\eth'+\bar{\tau}\right)\right]\debcc = 0\,.
\label{eq:DE}
\end{align}
The Debye potential then gives rise to the solution of Maxwell equations. For stationary axisymmetric fields we have
\begin{align}
\phi_0 &= \frac{1}{2}\frac{\p^2\,\debcc}{\p r^2}\,, \label{eq:phi0}\\
\phi_1 &= \frac{1}{2\sin\theta}\frac{\p^2}{\p r\p\theta}\left(\frac{\sin\theta\, \debcc}{\vrho}\right)
 -i\frac{\ka\sin\theta}{\vrho^3}\,\debcc\,, \\
\phi_2 &= -\frac{\Delta}{\vrho^2}\,\phi_0\,, \label{eq:phi2}
\end{align}
where Eq. (\ref{eq:phi2}) results from axisymmetry and stationarity.

Let us just shortly comment on another possibilities in choosing the Debye potential. Using the Debye potential with GHP weights $[0,0]$ does not lead to the Laplace equation and thus is not suitable for our purposes. Whereas by using the one with GHP weights $[0,2]$ under the assumptions of stationarity and axisymmetry does not lead to anything new. We can prove that if $\debcc_{[0,-2]}$ solves the Debye equation (\ref{eq:DE}), then $\bar{\chi}_{[2,0]}=-\bar{\vrho}^2\Delta^{-1}\debcc_{[0,-2]}$ solves the corresponding equation for this Debye potential and, moreover, it gives rise to exactly the same field.

\section{Generalized Axially Symmetrical Potential theory}
\label{sec:GASP}
It is straightforward to get the Green function of TME for $\phi_0$, $\phi_2$ since TME reduces to Laplace equation in a fiducial flat space of dimension $2s+3$ (where $s$ is the spin weight of the particular NP field component) under the assumptions of axial symmetry and stationarity. This has been done by Linet in \cite{Linet1977,Linet1979}. The generalized Laplace equation is
\begin{equation}
    \Delta_s g = 
    \frac{1}{\rho}\,\delta\left(\rho-\rho_0\right) \delta\left(z-z_0\right)\,,
\label{eq:GASP-Laplace}
\end{equation}
where the Laplace operator $\Delta_s$ is defined as
\begin{equation}
\Delta_s \equiv \frac{\p^2}{\p z^2}+\frac{\p^2}{\p \rho^2}+\frac{1+2s}{\rho}\frac{\p}{\p \rho} \,.
\end{equation}
Using GASP, Linet has provided the Green function of Eq. (\ref{eq:GASP-Laplace}) in term of the integral for general $s$. In our case, when $s=1$, we have
\begin{equation}
g = \frac{\rho_0^2}{2\pi}\int_0^\pi \frac{\sin^2\alpha}{\left(\rho^2-2\rho\rho_0\cos\alpha +\rho_0^2+(z-z_0)^2\right)^{\nicefrac{3}{2}}}\; \d\alpha\,.
\label{eq:Green1}
\end{equation}

The Debye equation (\ref{eq:DE}) can also be transformed to the Laplace equation. Yet, for DE we no longer seek for Green function. We need to find the Debye potential (which we call superpotential in this case and denote $\Psi$) of this Green function. It is given by twice integrating Eq. (\ref{eq:phi0}) with $\phi_0=G$.

Let us introduce function $\Xi_\textbf{r}$  which is the Debye superpotential rescaled and cast in Weyl coordinates
\begin{equation}
\Psi_\textbf{r}(r,\theta) = \sin\theta\, \Delta(r)\,\Xi_\textbf{r}(\rho,z)\,.
\end{equation}
The Debye equation transformed to the Weyl coordinates takes the form of generalized Laplace equation
\begin{equation}
\Delta_1\, \Xi_\textbf{r} = 0 \,.
\end{equation}

In the Weyl coordinates Linet \cite{Linet1979} obtained by simple integration values of the function $\Xi_\textbf{r}$ on the axis of the symmetry 
\begin{equation}
\Xi_\textbf{r}(0,z)=\frac{\pi}{\sin\theta_0}\frac{\sqrt{(z-\zz)^2+\rz^2}}{z^2-\beta^2}\,.
\label{eq:bc}
\end{equation}
A general theorem ensures that the solution of Laplace equation is in the axisymmetric case completely determined by its values on the axis. From GASP it thus follows that the superpotential is obtained by integration
\begin{multline}
    \Xi_\textbf{r}(\rho,z)
    =\frac{2}{\pi}\int_0^\pi \Xi_\textbf{r}\left(0,z+i\rho\cos\alpha\right)\sin^2\alpha\,\d\alpha \\
    =\frac{2}{\sin\theta_0} \int_0^\pi \frac{\sqrt{\rho_0^2+\left(z+i\rho\cos\alpha-\zz\right)^2}}{\left(z+i\rho\cos\alpha\right)^2-\beta^2}\,\sin^2\alpha\;\d\alpha \,.
\label{eq:integral}
\end{multline}

So far, the results of Linet.

\section{Exact integrals}
\label{sec:Int}
The integration of the Debye superpotential, Eq. (\ref{eq:integral}), is long, involves several steps of simplification and extensive use of identities involving elliptic integrals. Therefore we present only the results and discuss the properties of the solution.

Although the $\phi_0$ component does not carry any information about the monopole contribution of the central black hole, the Debye superpotential can contain a~monopole term -- it arises during integration as an integration constant. Thus, the proper value of the monopole on the central black hole has to be evaluated later.

Let us introduce 
\begin{align}
    h(z',\rho')&=z-z'+i\left(\rho-\rho'\right)\,,\\
    d(z',\rho')&=\sqrt{h(z',\rho')\bar{h}(z',\rho')} \nonumber\\
               &=\sqrt{\left(z-z'\right)^2+\left(\rho-\rho'\right)^2}\,;
\end{align}
we may think of $h$ as being a vector in complex plane connecting point $(z,\rho)$ and $(z',\rho')$, then $d$ is its norm.

The common form of elliptic modulus  for circular sources is
\begin{align}
    m &= \frac{4\rho\rz}{\left(z-\zz\right)^2+\left(\rho+\rz\right)^2}\,.
\end{align}
However our results will be given also in terms of complementary modulus $m'=1-m$ and reciprocal complementary modulus $\mu'=1/m'$, explicitly
\begin{align}
\mu'(\rz)&=1+\frac{4\rho\rz}{\left(z-\zz\right)^2+\left(\rho-\rz\right)^2}\,, \\
m'(\rz)&=\mu'(-\rz)\,.
\end{align}

Let us express our desired solution $\Xi_\textbf{r}$ of the Laplace's equation in terms of an auxiliary function $f$ which is defined as follows
\begin{widetext}
\begin{align}
    f(\rz)&=\frac{1}{\rho^2 d(\zz,\rz)}\Biggl[
    -i d(\zz,\rz)^2 E(\mu')+2\rz\left(4z-h(\zz,\rz)\right) K(\mu') 
    -4(z+\zz)\rz \Pi\left(\frac{h(\zz,-\rz)}{h(\zz,\rz)}\Big|\mu'\right) \nonumber \\
    &\qquad
    +\frac{2\rz d(\beta,0)^2}{\beta}\,
    \Pi\left(\frac{(\zz-\beta-i\rz)\bar{h}(\zz,-\rz)}
                  {(\zz-\beta+i\rz)\bar{h}(\zz,\rz)}\Big|\mu'\right) 
    -\frac{2\rz d(-\beta,0)^2}{\beta}\,
    \Pi\left(\frac{(\zz+\beta-i\rz)\bar{h}(\zz,-\rz)}
                  {(\zz+\beta+i\rz)\bar{h}(\zz,\rz)}\Big|\mu'\right)\Biggr]\,.
\label{eq:f}
\end{align}
\end{widetext}

\begin{figure}
\begin{center}
\subfloat[$-\beta<\zz<\beta,\beta=1$]{\includegraphics[width=.22\textwidth,keepaspectratio]{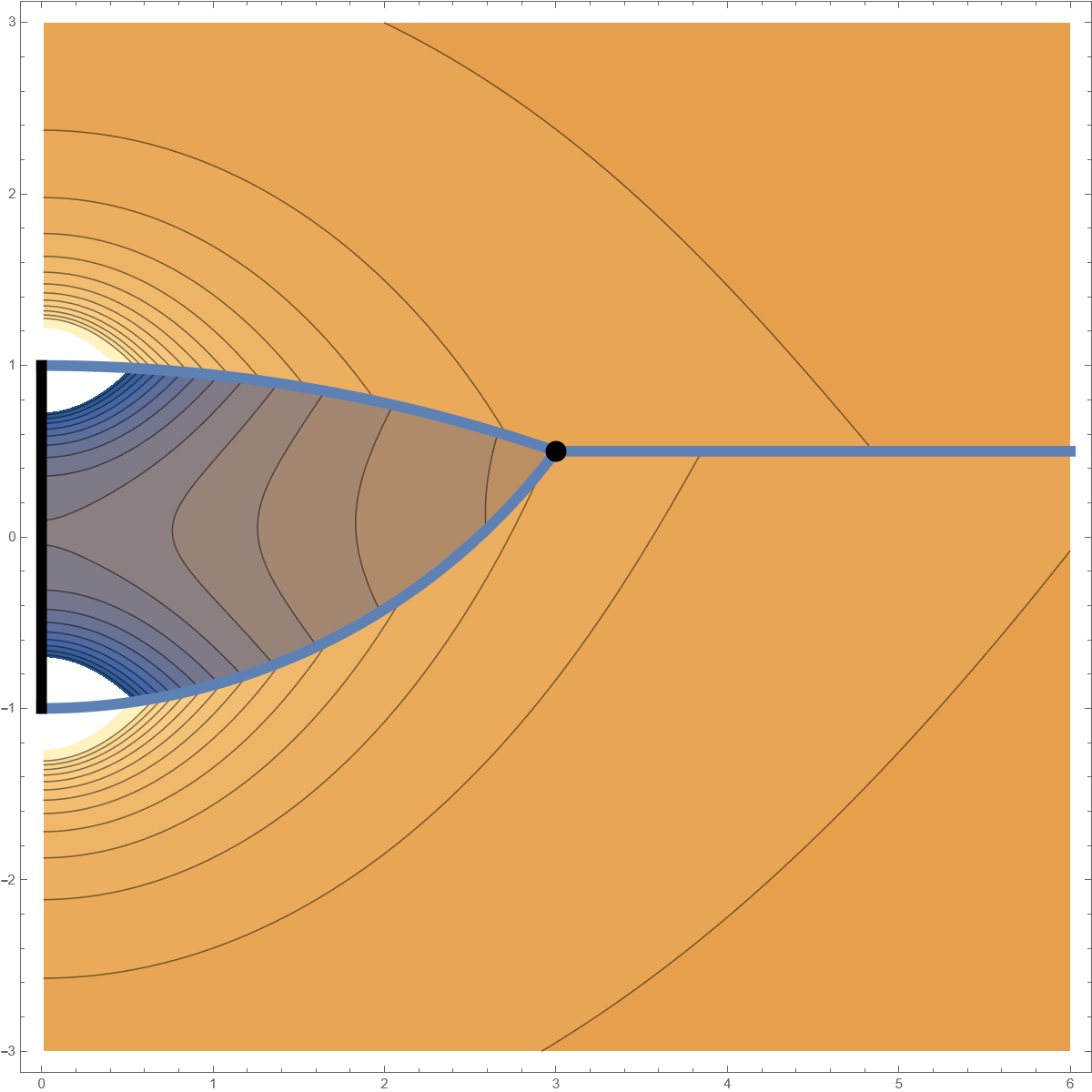}}\qquad
\subfloat[$\zz>\beta,\beta=1$]{\includegraphics[width=.22\textwidth,keepaspectratio]{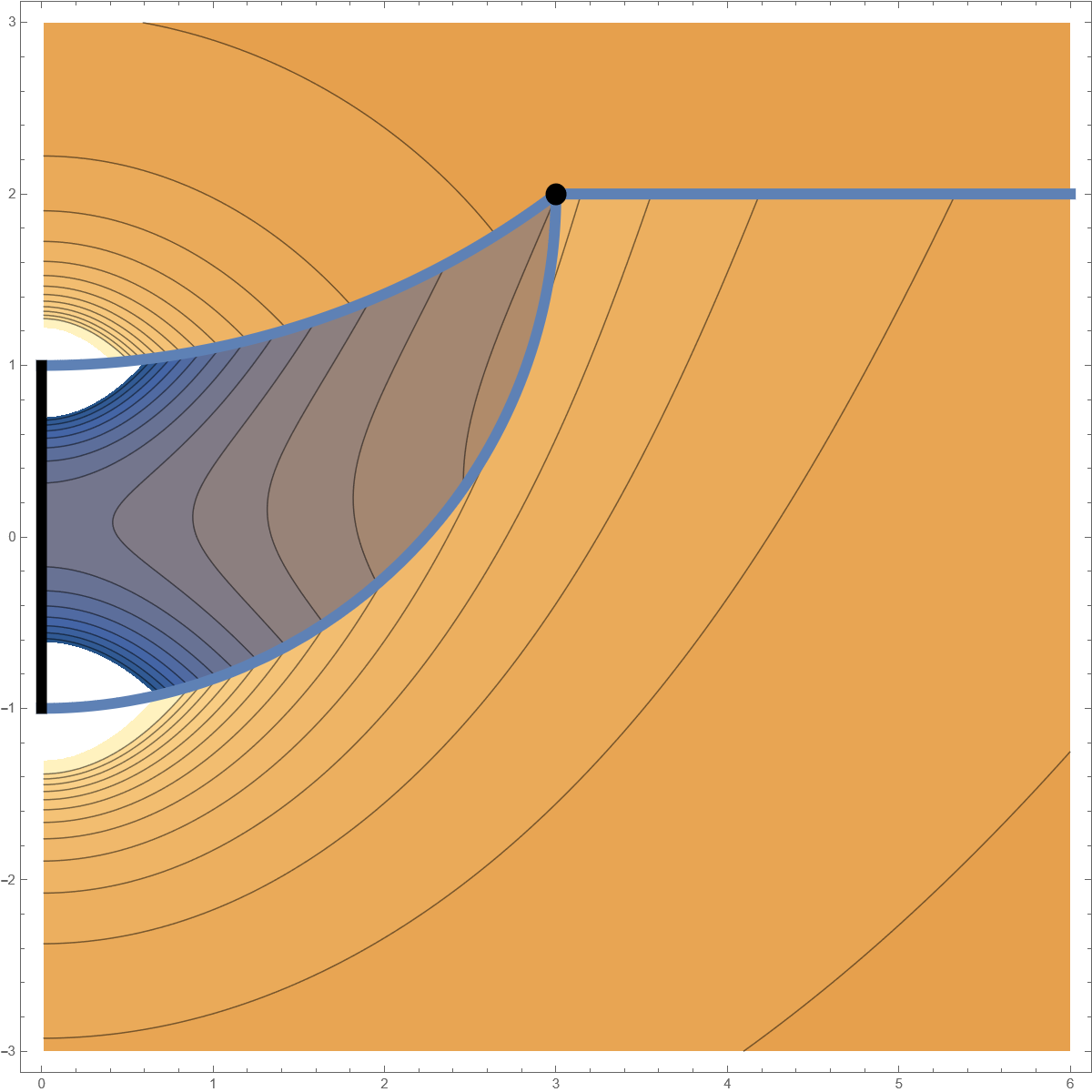}}
\caption{The contourplot of the Debye superpotential $\Xi_\textbf{r}$ in the Weyl coordinates ($\rho,z$). Discontinuities are present along the thick blue lines: $\vec{\gamma}_\textbf{n},\,\vec{\gamma}_\textbf{i},\,\vec{\gamma}_\textbf{s}$ and they divide the space into three different regions whose characteristic function are $\Theta_\textbf{n},\,\Theta_\textbf{i},\,\Theta_\textbf{s}$ (northern, inner and southern region). The outer horizon of the black hole stretches on the $z$ axis from $-\beta$ to $\beta$.}
\label{fig:jumps1}
\end{center}
\end{figure}

Then, the Debye superpotential for the circular sources $\Xi_\textbf{r}$ reads (mirror symmetry $\overbar{\Pi(n,m)}=\Pi(\bar{n},\bar{m})$ is used)
\begin{equation}
\Xi_\textbf{r} = \frac{1}{\sin\theta_0}\left(f(\rz)+\bar{f}(-\rz)\right) \,.
\label{eq:gD}
\end{equation}
It is clearly a \emph{real} function and has interesting structure of discontinuities as can be seen from the contourplot in Fig. \ref{fig:jumps1}. Two of these three discontinuities will be dealt with soon.

The existence of these discontinuities arise naturally from the behavior of elliptic integrals of the third kind $\Pi(n,m)$. Seen as a function of complex $n$ it has a branch cut on the interval $(1,\infty)$. When the elliptic characteristic $n$ crosses the real line for $n>1$ it thus has a step
\begin{align}
\lim_{\epsilon\rightarrow +0} \Pi(n-i\epsilon|m) &= \Pi(n|m)\,, \\
\lim_{\epsilon\rightarrow +0} \Pi(n+i\epsilon|m) &= \Pi(n|m)+\frac{\pi}{\sqrt{1-n}\sqrt{1-\frac{m}{n}}}\,. \label{eq:stepPi}
\end{align}

In our case, the discontinuities are located -- in the Weyl coordinates -- at two arcs connecting the north pole and south pole of the black hole with the source (curves $\vec{\gamma}_\textbf{n}$ and $\vec{\gamma}_\textbf{s}$) and a line from the source to infinity (curve~$\vec{\gamma}_\textbf{i}$).

The respective circles have centers $(0,z_\textbf{j})$ and radii $R_\textbf{j}$ where
\begin{equation}
\begin{aligned}
z_\textbf{n} &= \frac{1}{2}\frac{\zz^2+\rz^2-\beta^2}{\zz-\beta} \,, &
R_\textbf{n} &= \frac{1}{2}\frac{(\zz-\beta)^2+\rz^2}{\zz-\beta} \,, \\
z_\textbf{s} &= \frac{1}{2}\frac{\zz^2+\rz^2-\beta^2}{\zz+\beta} \,, &
R_\textbf{s} &= \frac{1}{2}\frac{(\zz+\beta)^2+\rz^2}{\zz+\beta} \,. 
\end{aligned}
\end{equation}
We also define
\begin{equation}
\begin{aligned}
r_\textbf{n} &= \rho^2+(z-z_\textbf{n})^2-R_\textbf{n}^2 \,, \\
r_\textbf{s} &= \rho^2+(z-z_\textbf{s})^2-R_\textbf{s}^2\,,
\end{aligned}
\end{equation}
for the purpose of definition of region functions.

\begin{figure}
\begin{center}
\subfloat[]{\includegraphics[keepaspectratio,height=.22\textwidth]{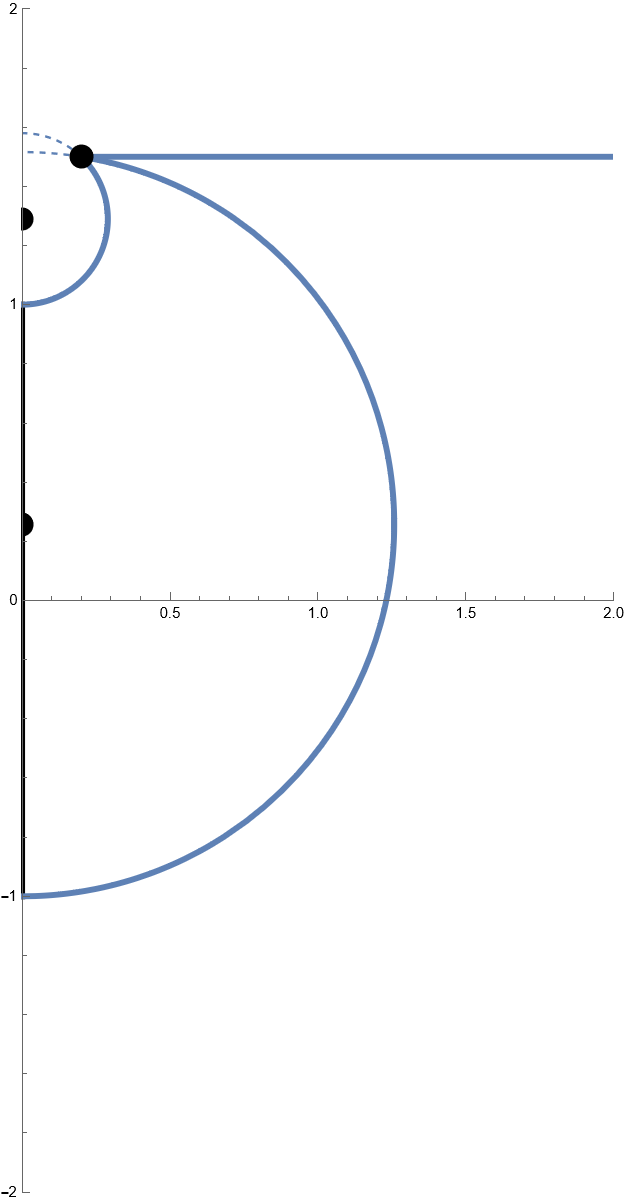}}\
\subfloat[]{\includegraphics[keepaspectratio,height=.22\textwidth]{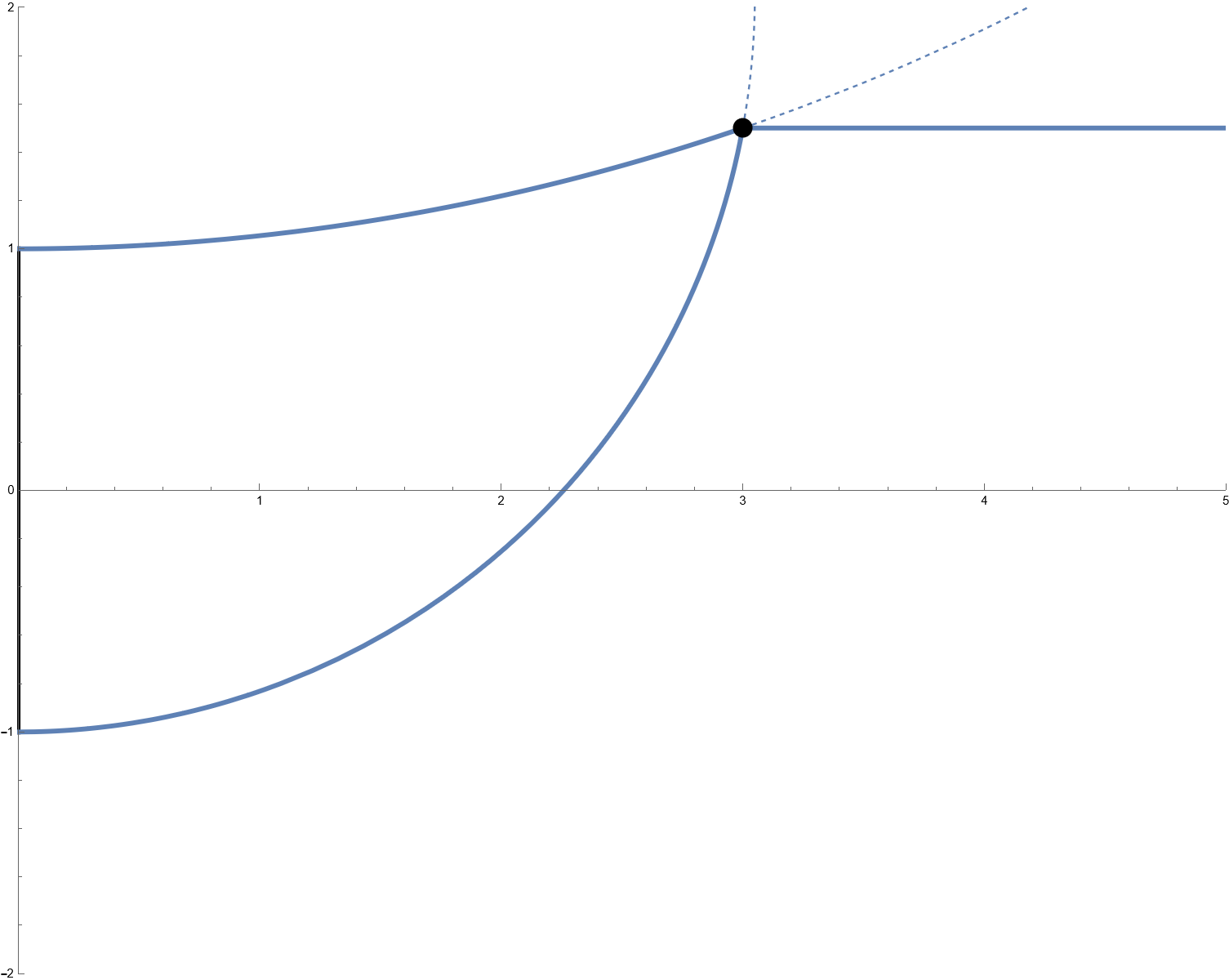}}\
\subfloat[]{\includegraphics[keepaspectratio,height=.22\textwidth]{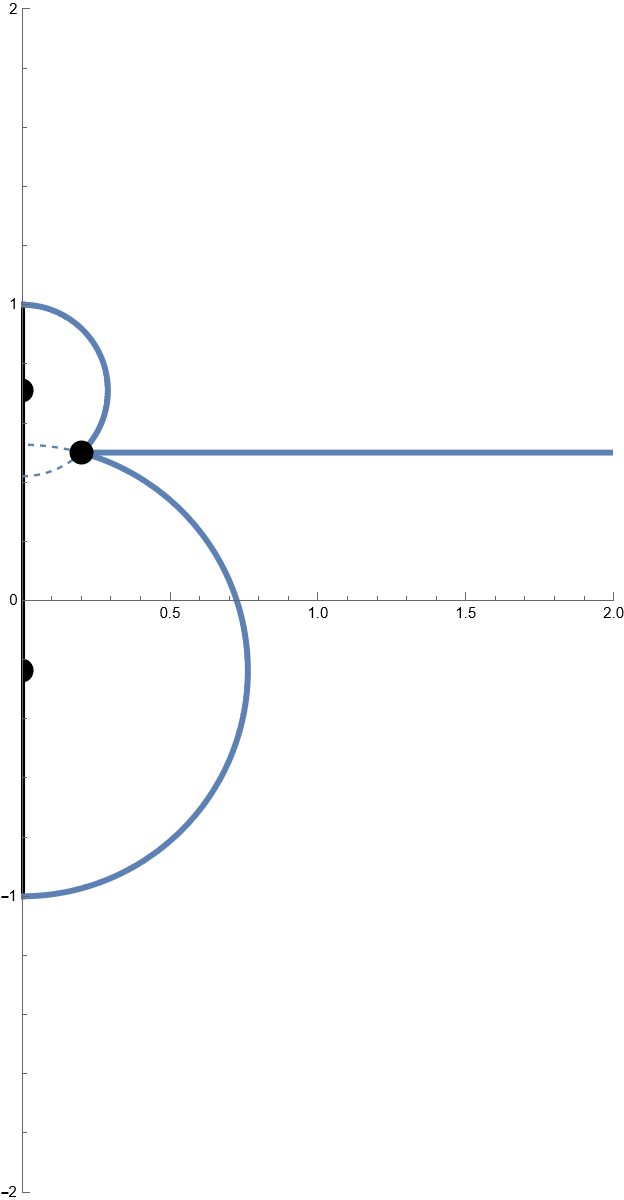}}\
\subfloat[]{\includegraphics[keepaspectratio,height=.22\textwidth]{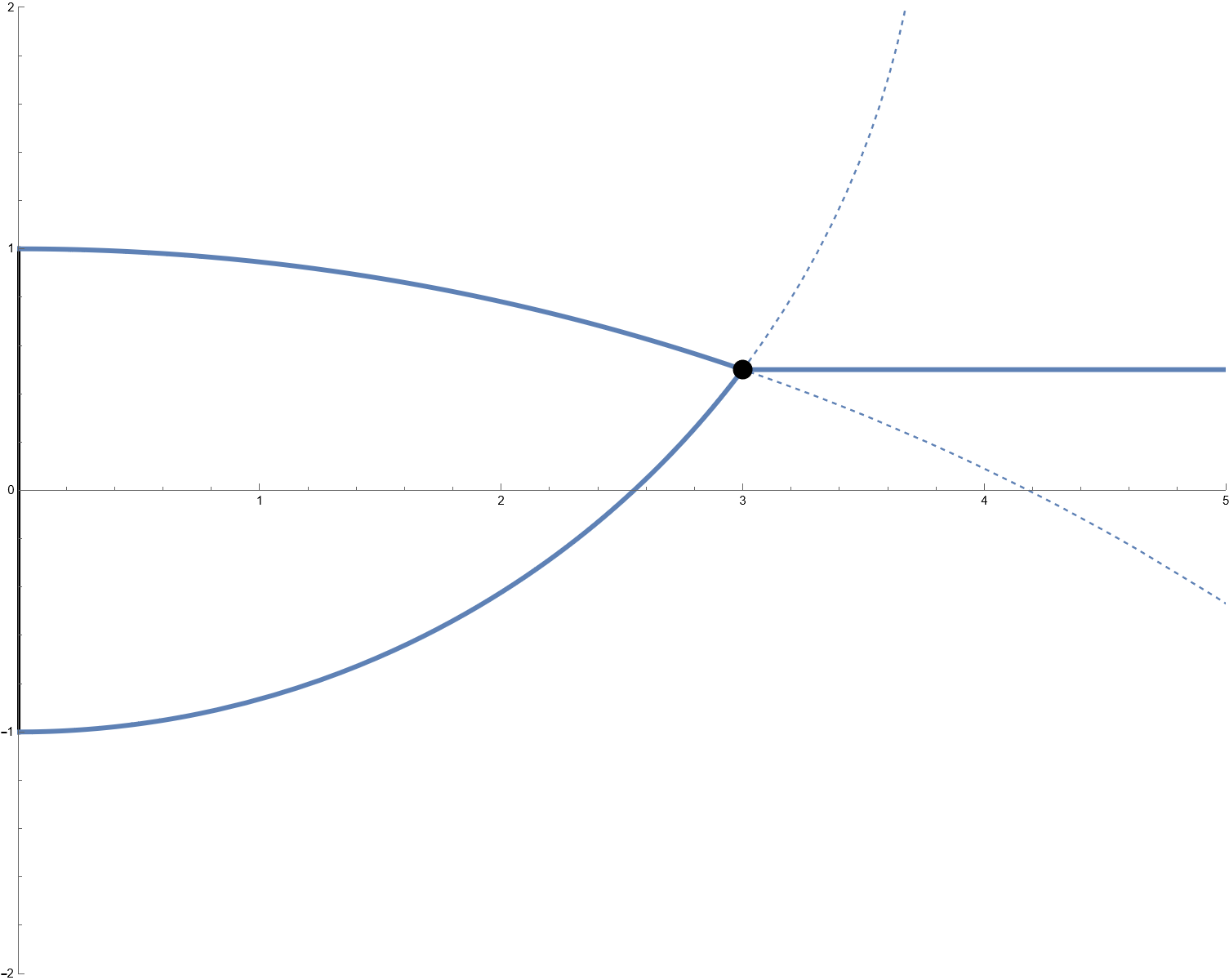}}
\caption{Diagram showing different possibilities of the location of discontinuities depending on the mutual position of the horizon and the source in Weyl coordinates. The shape of inner region has nontrivial algebraic expression. The black hole horizon stretches on vertical axis from $-1$ to $1$ and the location of the ring is denoted by a point. Wherever possible the centers of the circles are also shown (dots on the axis).}
\end{center}
\end{figure}

The Weyl plane is divided into the north, inner and south region with the region functions defined as
\begin{align}
\Theta_\textbf{n} &= \Theta(+z-z0) \nonumber\\
                  &\  +\sign(\zz-\beta)\Theta\left(-r_\textbf{n}\right)
                     \Theta\left[-(z-\zz)\sign(\zz-\beta)\right]\,, \nonumber\\
\Theta_\textbf{i} &= 
\begin{cases}
\Theta(-\sign(z_\textbf{s}+\beta)r_\textbf{s}) \Theta(\sign(z_\textbf{n}-\beta)r_\textbf{n})\,,\\
\qquad\qquad \text{for }|z_\textbf{n}|>\beta \text{ or }|z_\textbf{s}|>\beta \, \vspace{1em} \\ 
\Theta(-r_\textbf{n})+\Theta(-r_\textbf{s})\\
-\Theta(-\sign(z_\textbf{s}+\beta)r_\textbf{s}) \Theta(\sign(z_\textbf{n}-\beta)r_\textbf{n})\,, \\
\qquad\qquad \text{otherwise}
\end{cases}\hspace{-1em} \\
\Theta_\textbf{s} &= \Theta(-z+z0) \nonumber \\
                  &\  -\sign\left(\zz+\beta\right)\Theta\left(-r_\textbf{s}\right)
                    \Theta\left[-(z-\zz)\sign(\zz+\beta)\right] \,, \nonumber
\end{align}
where $\Theta(x)$ stands for Heaviside step function.

We have realized that the discontinuities across the lines $\vec{\gamma}_{\textbf{n}},\vec{\gamma}_{\textbf{i}},\vec{\gamma}_{\textbf{s}}$ corresponds to a contribution of the Debye potential of monopole in one part and zero in the other in the sense that 
\begin{equation}
\left[\Xi_\textbf{r}\right]|_{\vec{\gamma}_\textbf{j}}=\Xi_\textbf{j}|_{\vec{\gamma}_\textbf{j}}\,, 
\end{equation}
for $\textbf{j}\in\{\textbf{n},\,\textbf{i},\,\textbf{s}\}$, where $[f(x)]$ represents the jump. Thus we may get rid of the discontinuities across the lines $\vec{\gamma}_\textbf{n}$ and $\vec{\gamma}_\textbf{s}$ by adding an appropriate monopole term in the respective regions as
\begin{equation}
\begin{aligned}
\Xi &= \Xi_\textbf{r}
-\frac{4i\pi}{\sin\theta_0} \frac{\left(\beta+i\ka\right)\sqrt{(\zz-\beta)^2+\rz^2}}{\beta}\;
\Xi_\textbf{n}\Theta_\textbf{n} \\
&\phantom{=\Xi_\textbf{r}}+\frac{4i\pi}{\sin\theta_0} \frac{\left(\beta-i\ka\right)\sqrt{(\zz+\beta)^2+\rz^2}}{\beta}\;
\Xi_\textbf{s}\Theta_\textbf{s}\,,
\label{eq:XiSmooth}
\end{aligned}
\end{equation}
where the normalized --- corresponding to unit charge --- Debye potentials of the monopole read
\begin{align}
\Xi_\textbf{n}&=-i\frac{\sqrt{(z-\beta)^2+\rho^2}}{{2\left(\beta+i\ka\right)\rho^2}}\,,\\
\Xi_\textbf{i}&=-i\frac{z+\zz}{(r_p+r_m)\rho^2}\,,\\
\Xi_\textbf{s}&=+i\frac{\sqrt{(z+\beta)^2+\rho^2}}{{2\left(\beta-i\ka\right)\rho^2}}\,.
\end{align}
The Debye superpotential $\Xi$ remains \emph{real}.

Actually the discontinuity can be removed across arbitrary two of these three lines, yet it has to remain present on the third one. It has to be stressed that it is necessary to remove two of these three discontinuities -- if this is not done, then the electromagnetic field component $\phi_1$ generated from this superpotential is discontinuous (due to the presence of different monopole contributions). 

The remaining discontinuity is caused by a ramification of a multi-valued function. Yet, it can be also seen as a presence of distributional sources on the right hand side of the Laplace's equation, and we have decided to have these sources along $\vec{\gamma}_\textbf{i}$. 

The function $\Xi$ is finally sufficiently smooth across $\vec{\gamma}_\textbf{n}$ and $\vec{\gamma}_\textbf{s}$; but the discontinuity across $\vec{\gamma}_\textbf{i}$ is still present. 

\begin{figure}
\begin{center}
\subfloat[]{\includegraphics[width=.22\textwidth,keepaspectratio]{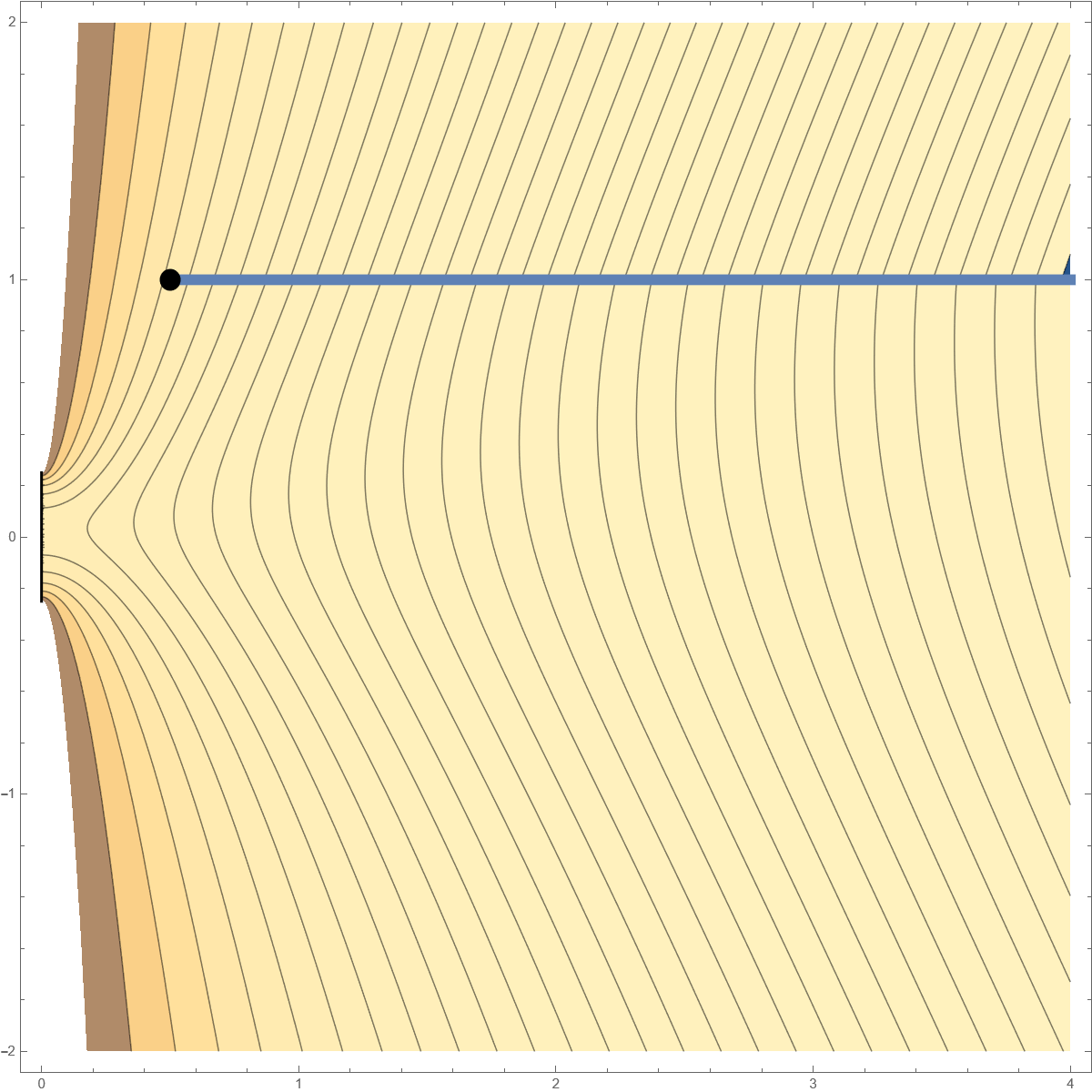}}\qquad
\subfloat[]{\includegraphics[width=.22\textwidth]{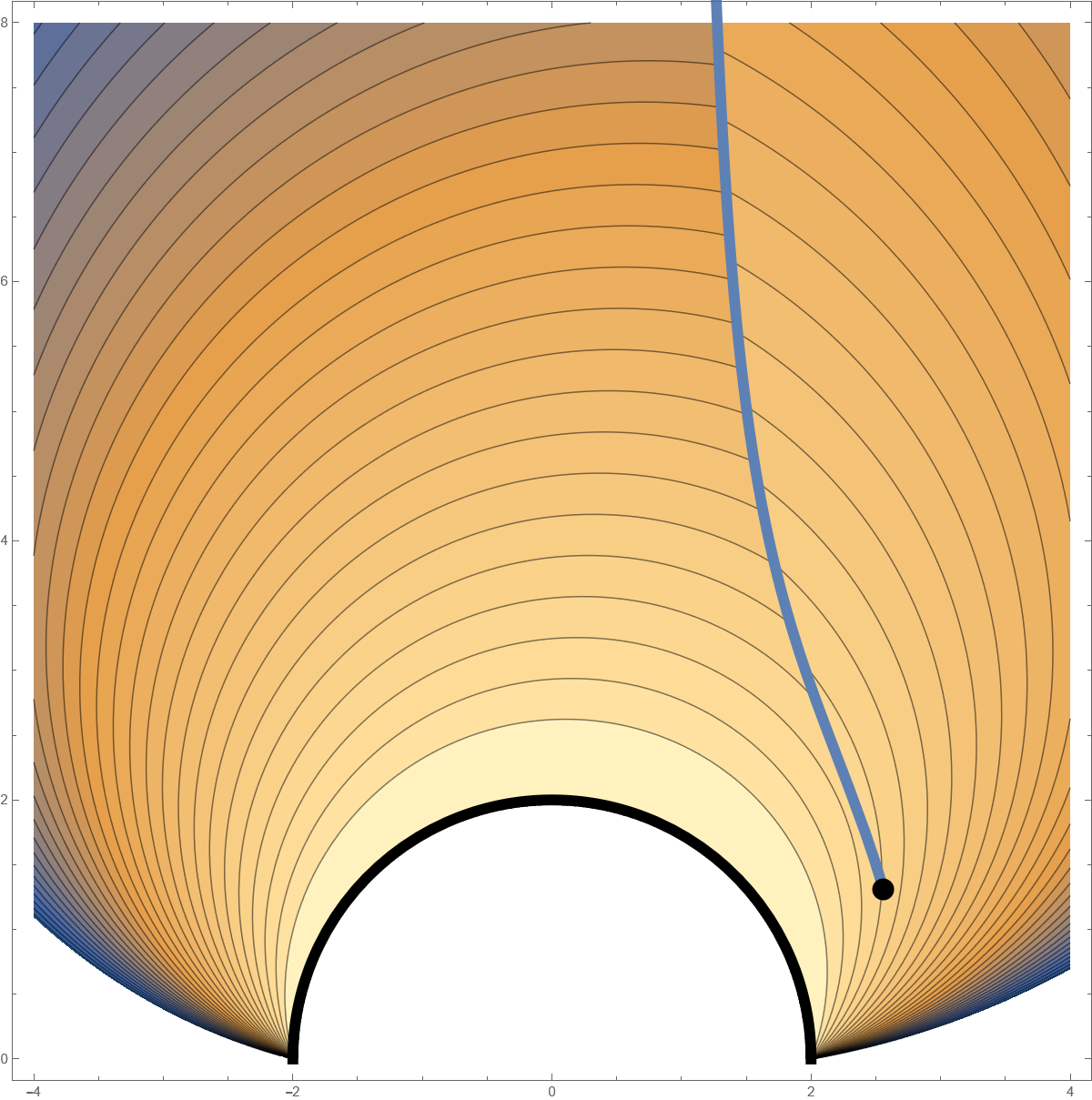}}
\caption{The contourplot of the Debye superpotential (a) $\Xi$ in Weyl coordinates ($\rho,z$) and (b) $\Psi$ in Boyer\,--\,Lindquist coordinates $(r,\theta)$. Discontinuity is still present along the line $\vec{\gamma}_\textbf{i}$ --- thick blue line. The white regions are merely a cut-off of the values.}
\label{fig:jumps}
\end{center}
\end{figure}

The Debye superpotential is
\begin{equation}
\Psi = \sin\theta\, \Delta(r)\; \Xi(r,\theta)\,,
\end{equation}
where $\Xi(\rho,z)$ given by Eq. (\ref{eq:XiSmooth}) has to be transformed from Weyl coordinates to Boyer\,--\,Lindquist coordinates. The contourplot of $\Xi$ and $\Psi$ are in Fig. \ref{fig:jumps}.

For stationary axisymmetric sources we may write
\begin{equation}
\vec{J}=j_0 \vec{\p_t}+j_3\vec{\p_\phi} \,,
\label{eq:vecJ}
\end{equation}  
where $j_0$ and $j_3$ are functions of $r$ and $\theta$ only.

Simplified expression for the sources of TME reads as follows
\begin{multline}
J_0 = \frac{1}{2\vrho\Sigma}\Biggl[
    -\frac{\p}{\p\theta}\left(\vrho^2 j_0\right)+i\ka\sin\theta\frac{\p}{\p r}\left(\vrho^2 j_0\right) \\
    \frac{\p}{\p\theta}\left(\ka\sin^2\theta\,\vrho^2\, j_3\right)-i\frac{\p}{\p r}\biggl(\left(\ka^2+r^2\right)\sin\theta\,\vrho^2\, j_3\biggr)\Biggr]
\end{multline}
and the Debye potential is given by convolution of the Debye superpotential with sources as
\begin{multline}
\debcc = \int_0^\pi \int_0^\infty \Psi(r,\theta,r',\theta')J_0(r',\theta',r_0,\theta_0)\\ \Sigma(r',\theta')\sin\theta'\,\d r'\d\theta' \,.
\label{eq:convolution2}
\end{multline}

We can also explicitly integrate the Green function $G$ of the Teukolsky operator which is 
\begin{align}
G&=\frac{\sin\theta}{\sin\theta_0}\; g\,,\\
g&=\frac{d(\zz,-\rz)}{\rho^2}\left[-E(m)+\left(1+\frac{2\rz\rho}{d(\zz,-\rz)^2}\right)K(m)\right]\,.
\end{align}
The function $G$ solves TME and for $g$ we have $\Delta_1 g=0$. It can be checked that 
\begin{equation}
G = \frac{1}{2}\frac{\p^2\, \Psi}{\p r^2} \,,
\end{equation}
which is a consistency check following from the definition of the superpotential.

Let us also note that the position of discontinuities discussed so far is ``natural'' in the sense, that it is defined by the branch cuts of respective elliptic integrals of the third kind. These discontinuities are merely mathematical difficulties, see Appendix \ref{app:Riem} for details, the fields of realistic physical sources are well behaved as we will see later. 

\begin{figure}
\begin{center}
\subfloat[$\Xi_\textbf{0}$]{\includegraphics[width=.15\textwidth,keepaspectratio]{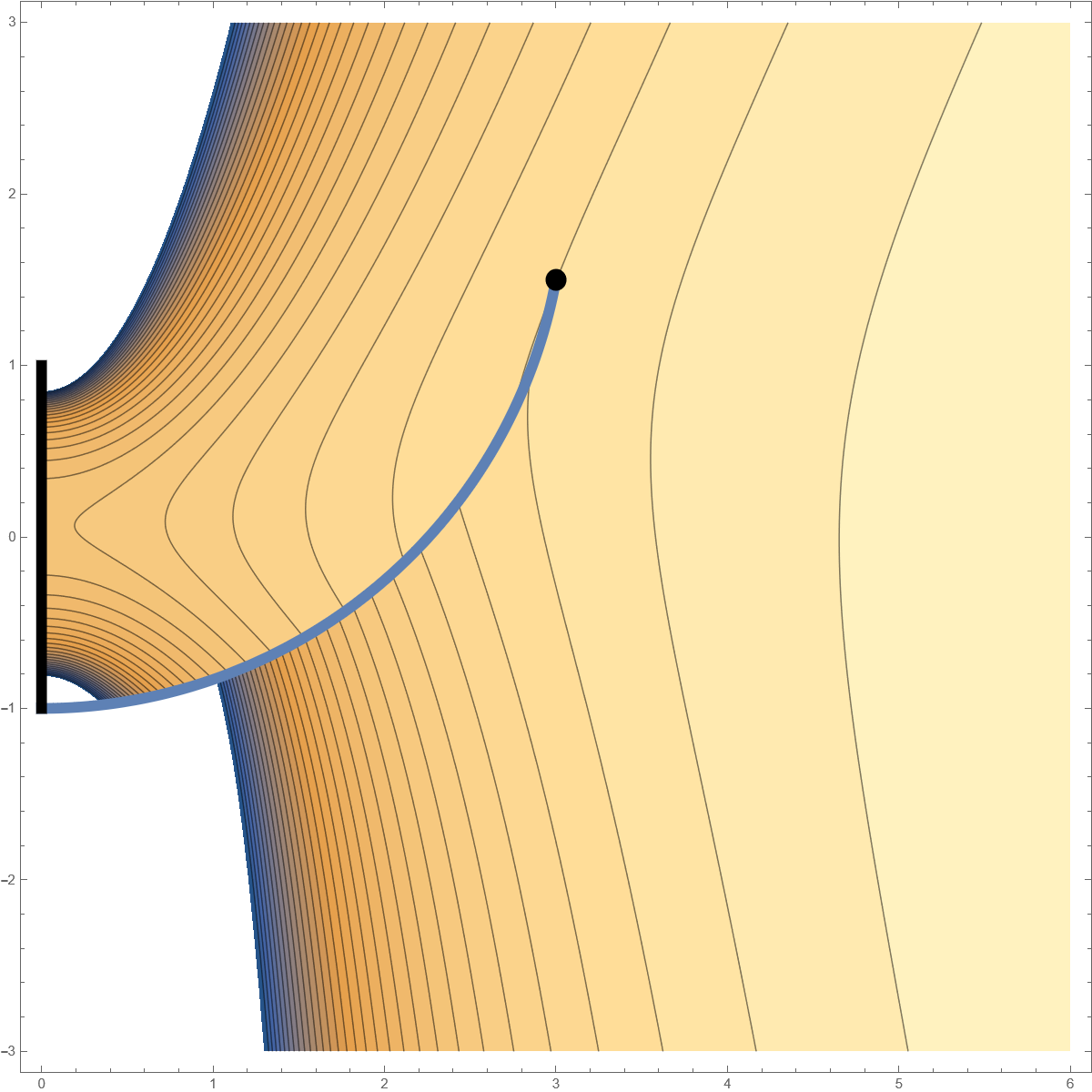}}\;
\subfloat[$\Xi_\textbf{1}$]{\includegraphics[width=.15\textwidth]{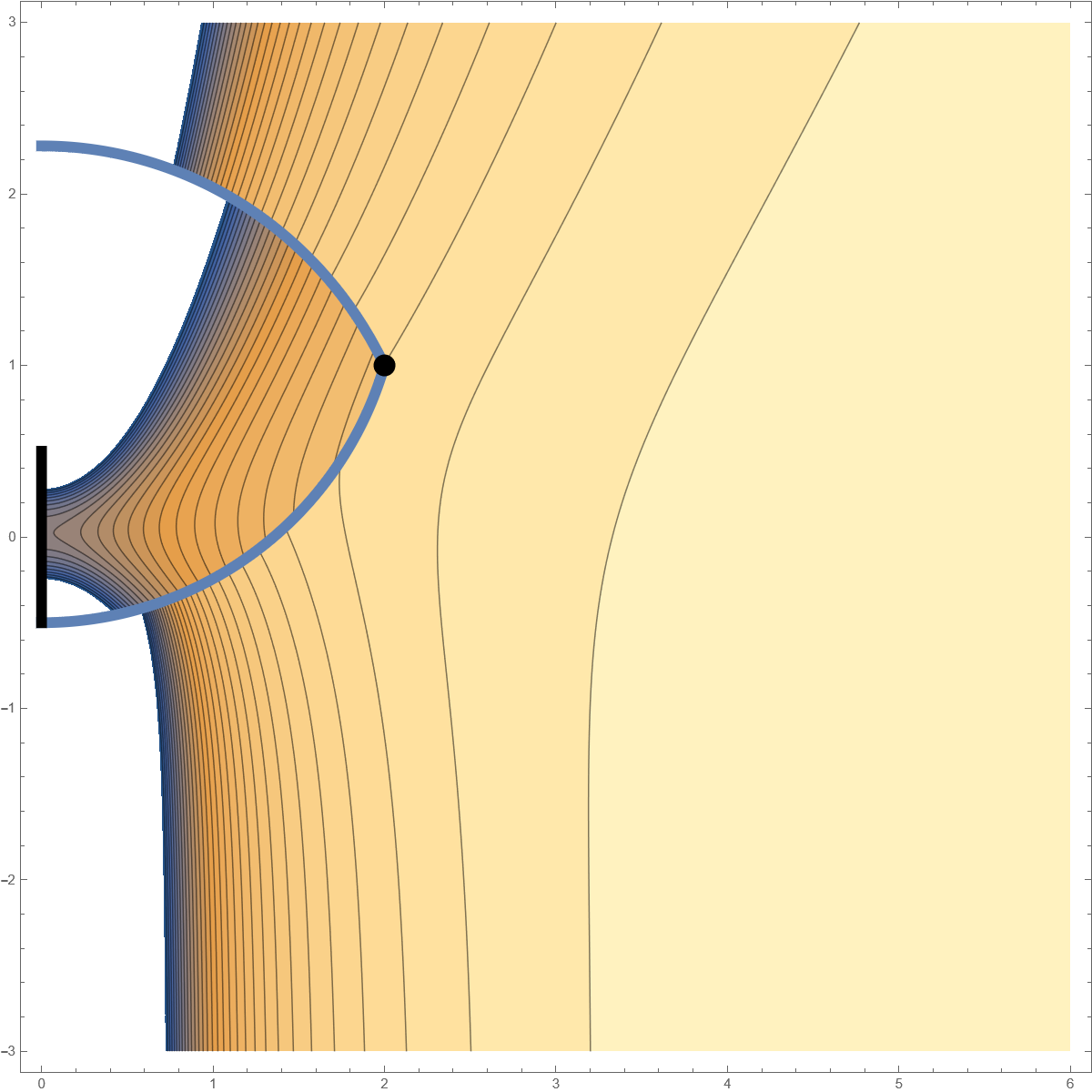}}\;
\subfloat[$\Xi_\textbf{2}$]{\includegraphics[width=.15\textwidth]{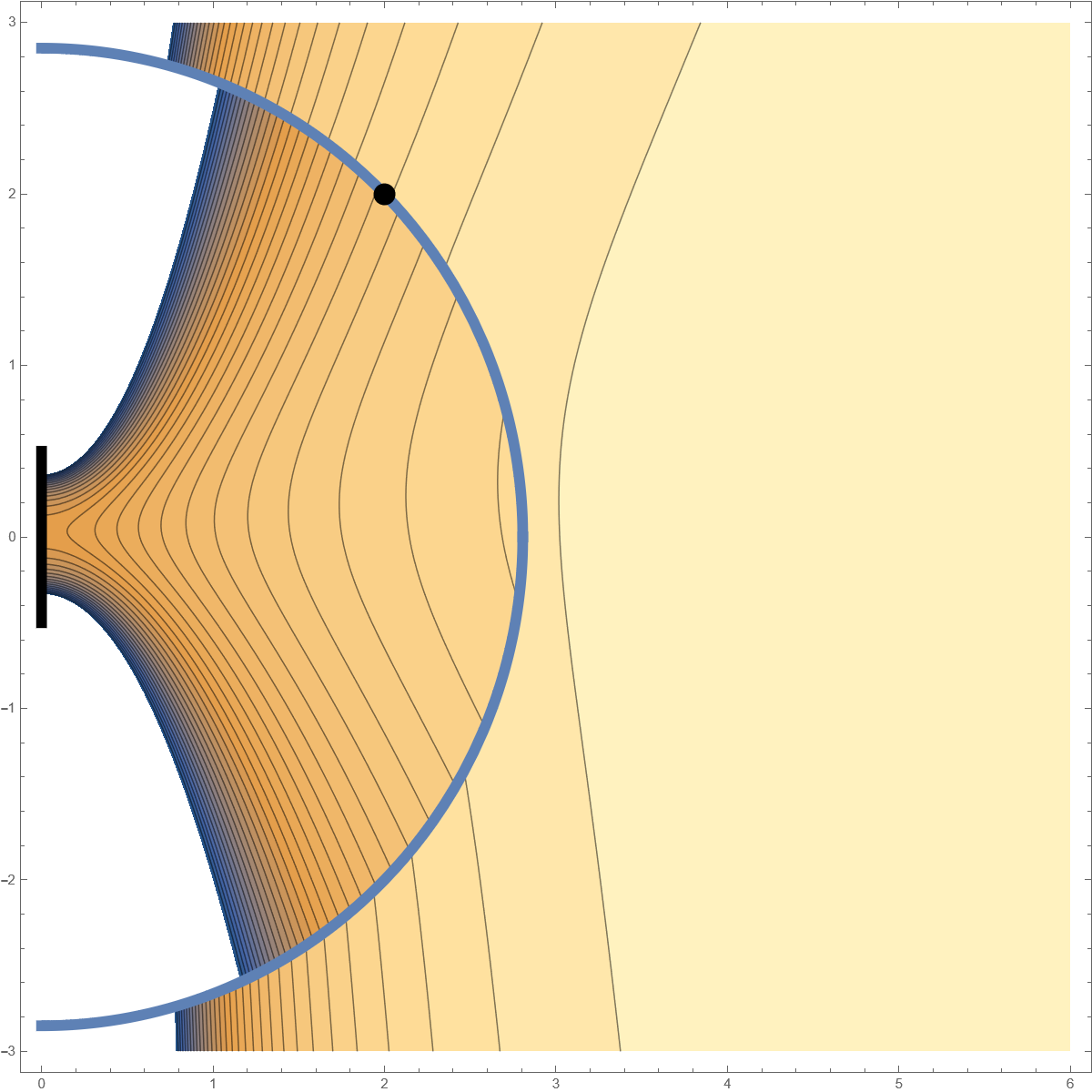}}
\caption{The contourplot of the Debye superpotential $\Xi_\textbf{j}$, $\textbf{j}\in (\textbf{0},\textbf{1},\textbf{2})$ in Weyl coordinates with different positions of discontinuities. In (c) we can see the discontinuities corresponding to the series expansion case. The exact formulae can be found in Appendix \ref{app:discont}.}
\label{fig:jumps2}
\end{center}
\end{figure}

But we are allowed to move these discontinuities wherever is desired by analytical continuation and taking a new branch cut. Thus, they can be moved to line $r=r_0$ on Boyer\,--\,Lindquist coordinates (which is an ellipse in Weyl coordinates). The reason we make this short comment is to draw a clear theoretical connection to the series expansion approach. In \cite{Bicak1976} the field is given by different series expansions in regions $r<r_0$ and $r>r_0$
\begin{align}
\phi_0 &= 2\sum_{l=1}^\infty\frac{a_l}{l(l+1)} {}_1Y_{l0} \, \frac{\d^2 y^{(1)}_{l0}}{\d x^2} \,, \nonumber\\ 
y^{(1)}_{l0} &= x(x-1) F\left(l+2,1-l,2;x\right) \,, &&\text{for }r<r_0\,,\\
\phi_0 &= 2\sum_{l=1}^\infty\frac{b_l}{l(l+1)} {}_1Y_{l0} \, \frac{\d^2 y^{(2)}_{l0}}{\d x^2} \,, \nonumber\\
y^{(2)}_{l0} &= (-x)^l F\left(l,1+l,2l+2;x\right)\,,&&\text{for }r>r_0\,,
\end{align}
where $x=\frac{r-r_m}{r_p-r_m}$. Thus, this is almost ready to be twice integrated along $r$ to obtain the series expansion of the Debye potential. As discussed in \cite{Bicak1976}, for $\phi_1$ a different monopole term has to be added in regions $r<r_0$ and $r>r_0$, so the discontinuity is present also in this formulation. In Fig. \ref{fig:jumps2} another possible locations of discontinuities are visualized. The respective formulae are postponed to Appendix \ref{app:discont}.

We do not integrate the series expansion of the Debye potential since we already have found an exact closed solution. For particular sources, we have numerically checked the validity of our results.

\subsection{The Debye superpotential}
\label{sec:superpotential}
The Debye superpotential is itself a Debye potential for some electromagnetic field. What will be the electric and magnetic  charge induced on the black hole? Recall that the charge within a topological sphere is given by the Gauss law -- Eq. (\ref{eq:totQ}). Using Eq. (\ref{eq:phi2}) this simplifies to
\begin{equation}
i Q_e-Q_m =  \int_0^\pi  
-i\sin\theta \left( r^2+\ka^2 \right) \phi_1  
+\frac{\ka\Delta\sin^2\theta}{\Krho}\,\phi_0
\ \d\theta \,,
\end{equation}
which we would like to evaluate on the horizon.

First of all we may express the electromagnetic field in terms of the Debye superpotential $\phi[\Psi_\textbf{r}]$. The behavior of $\Psi_\textbf{r}$ on the horizon\footnote{Keep in mind that during the ``regularization'' of the Debye super\-potential in Eq. (\ref{eq:XiSmooth}) no charge has been added on the black hole.} is of the form 
\begin{equation}
\Psi_\textbf{r}=0+S(\theta)(r-r_p)+O\left((r-r_p)^2\right).
\end{equation} 
In particular we have
\begin{multline}
\Psi_\textbf{r}\doteq 0 \\ 
-\pi\frac{\sqrt{4\Delta(r_0)\sin^2\theta_0+(-2\beta\cos\theta+\cos \theta_0\,\Delta'(r_0))^2}}{\beta\sin\theta\,\sin\theta_0}\,(r-r_p)\\+\ldots
\end{multline}
Evaluating the flux on the horizon and simplifying the expressions yield a simple result which can be explicitly integrated
\begin{multline}
i Q_e-Q_m = \int_0^\pi -ir_p(r_p+r_m)\frac{\p}{\p\theta}\left(\frac{\sin\theta\, \p_r \Psi_\textbf{r}}{\vrho(r_p,\theta)}\right) \,\d\theta \\
= -i r_p(r_p+r_m)\left[\frac{\sin\theta\, S(\theta)}{\vrho(r_p,\theta)}\right]_{\theta=0}^\pi \,.
\end{multline}
Thus, the total charge upon the black hole is\footnote{Notice, that $\vrho(r,0)\vrho(r,\pi)=r^2+r_pr_m$.}
\begin{multline}
Q_\textbf{r}=i Q_e-Q_m=-i\frac{\pi}{\beta\sin\theta_0}
\Biggl[\vrho(r_p,\theta) \times \\ \sqrt{4\Delta(r_0)\sin^2\theta_0+(-2\beta\cos\theta+\cos \theta_0\,\Delta'(r_0))^2}\Biggr]_{\theta=0}^\pi \,.
\label{eq:Qr}
\end{multline}

\subsection{Charged ring}
\label{sec:ring}
Let us consider the source of static charged ring in the form of Eq. (\ref{eq:vecJ}) with
\begin{align}
j_0 & = \hat{j}_0(r_0,\theta_0)\frac{\delta\left(r-r_0\right)}{\Sigma(r_0,\theta_0)}\frac{\delta\left(\theta-\theta_0\right)}{\sin\theta_0}\,, &
j_3 & =0\,.
\end{align}
Then, the convolution of the sources with the superpotential as in the Eq. (\ref{eq:convolution2}) leads to the Debye potential of charged ring
\begin{equation}
\debcc_{\textbf{ring}} = \frac{\hat{j}_0(r_0,\theta_0)}{2\overbar{\vrho(r_0,\theta_0)}}\left(\cot\theta_0 + \frac{\p}{\p\theta_0} - i\ka\sin\theta_0\,\frac{\p}{\p r_0}\right)\Psi \,.
\label{eq:opRing}
\end{equation}
From this Debye potential the electromagnetic field is easily reconstructed by differentiation. Hence, we have
\begin{align}
\phi_0 &= \phi_0[\debcc_{\textbf{ring}}] \,, \nonumber \\
\phi_1 &= \phi_1[\debcc_{\textbf{ring}}] + \frac{e_\textbf{ring}}{\vrho^2} \,, \\
\phi_2 &= \phi_2[\debcc_{\textbf{ring}}] \,, \nonumber
\end{align}
where the value of the charge $e_\textbf{ring}$ counterbalances the charge induced on the black hole by the presence of the ring. It is given by the same operator as in Eq. (\ref{eq:opRing}), i.e.
\begin{equation}
e_{\textbf{ring}} =  \frac{\hat{j}_0(r_0,\theta_0)}{2\overbar{\vrho(r_0,\theta_0)}}\left(\cot\theta_0 + \frac{\p}{\p\theta_0} - i\ka\sin\theta_0\,\frac{\p}{\p r_0}\right) Q_\textbf{r} \,.
\end{equation}

\begin{figure}
\begin{center}
\subfloat[Integral curves of $\vec{E}$.]{\includegraphics[width=.4\textwidth,keepaspectratio]{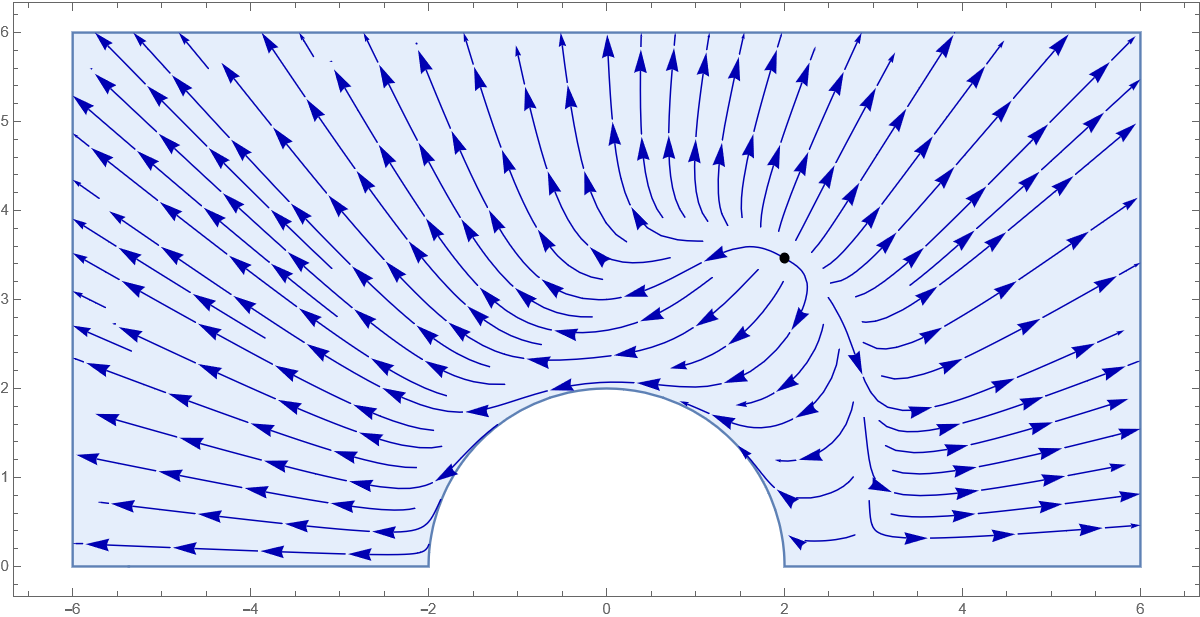}}\quad
\subfloat[Integral curves of $\vec{B}$.]{\includegraphics[width=.4\textwidth,keepaspectratio]{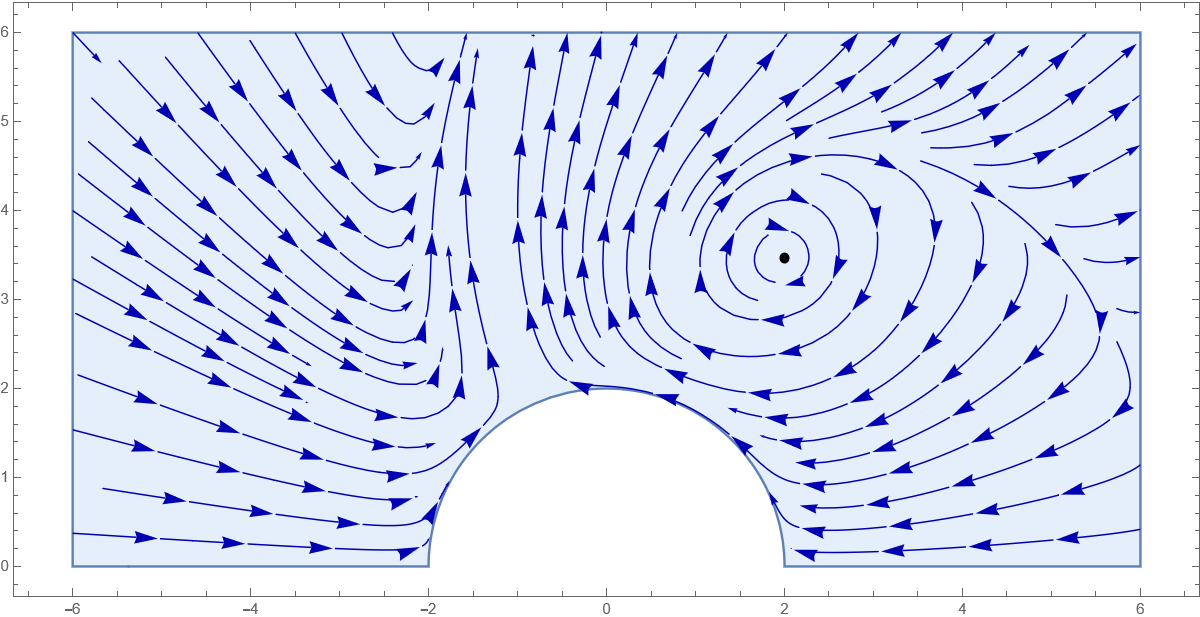}}
\caption{Integral curves of of the electric $\vec{E}$ and the magnetic $\vec{B}$ field of a charged ring (depicted by black dot) above the black hole as measured by ZAMO in the $(r,\theta)$ plane. The Meissner effect exhibits itself as well as the presence of both electric and magnetic field due to the almost extremal rotation of the black hole. The rotation axis is horizontal and the parameters are $r_p=2,r_m=1.999,r_0=4,\theta_0=\pi/3$.}
\label{fig:ring}
\end{center}
\end{figure}

The integral lines of electric and magnetic field of a charged ring hovering above the equatorial plane on the Kerr background which would have been measured by congruence of ZAMO observers are visualized in Fig. \ref{fig:ring}.

We have numerically compared the values of $\phi_0$ with the results given in \cite{Bicak1976} as an infinite series expansion and the results are identical (modulo normalization factor $\sqrt{2}\pi$).

\subsection{Current loop}
\label{sec:loop}
Let the source of the electromagnetic field be an axially symmetric current loop defined as 
\begin{align}
j_0 & =0\,, &
j_3 & = \hat{j}_3(r_0,\theta_0)\frac{\delta\left(r-r_0\right)}{\Sigma(r_0,\theta_0)}\frac{\delta\left(\theta-\theta_0\right)}{\sin\theta_0}\,. &
\end{align}
Evaluating the Eq. (\ref{eq:convolution}) leads to the Debye potential of the current loop
\begin{multline}
\debcc_{\textbf{current}} = \hat{j}_3(r_0,\theta_0) \frac{\sin\theta_0}{2\overbar{\vrho(r_0,\theta_0)}} \\
\left(-i r_0 -\ka\sin\theta_0 \frac{\p}{\p\theta_0} + i\left(r_0^2+\ka^2\right)\frac{\p}{\p r_0}\right)\Psi \,.
\end{multline}
Again, the results are in agreement with \cite{Bicak1976} if we set the normalization constant $\hat{j}_3=2\sqrt{2}r_0\sqrt{\Delta(r_0)/\Upsilon(r_0,\pi/2)}$ (in \cite{Bicak1976} the ring is only in equatorial plane).

The field can be reconstructed from the NP projections 
\begin{align}
\phi_0 &= \phi_0[\debcc_{\textbf{current}}] \,, \nonumber \\
\phi_1 &= \phi_1[\debcc_{\textbf{current}}]+\frac{e_{\textbf{current}}}{\vrho^2}\,, \\
\phi_2 &= \phi_2[\debcc_{\textbf{current}}] \,, \nonumber
\end{align}
where the monopole charge $e_{\textbf{current}}$ has to be set to
\begin{multline}
e_{\textbf{current}} = \hat{j}_3(r_0,\theta_0) \frac{\sin\theta_0}{2\overbar{\vrho(r_0,\theta_0)}}\\ 
\left(-i r_0 -\ka\sin\theta_0 \frac{\p}{\p\theta_0} + i\left(r_0^2+\ka^2\right)\frac{\p}{\p r_0}\right)Q_\textbf{r} \,,
\end{multline}
if we want the black hole to be uncharged.

\begin{figure}
\begin{center}
\subfloat[Integral curves of $\vec{E}$.]{\includegraphics[width=.4\textwidth,keepaspectratio]{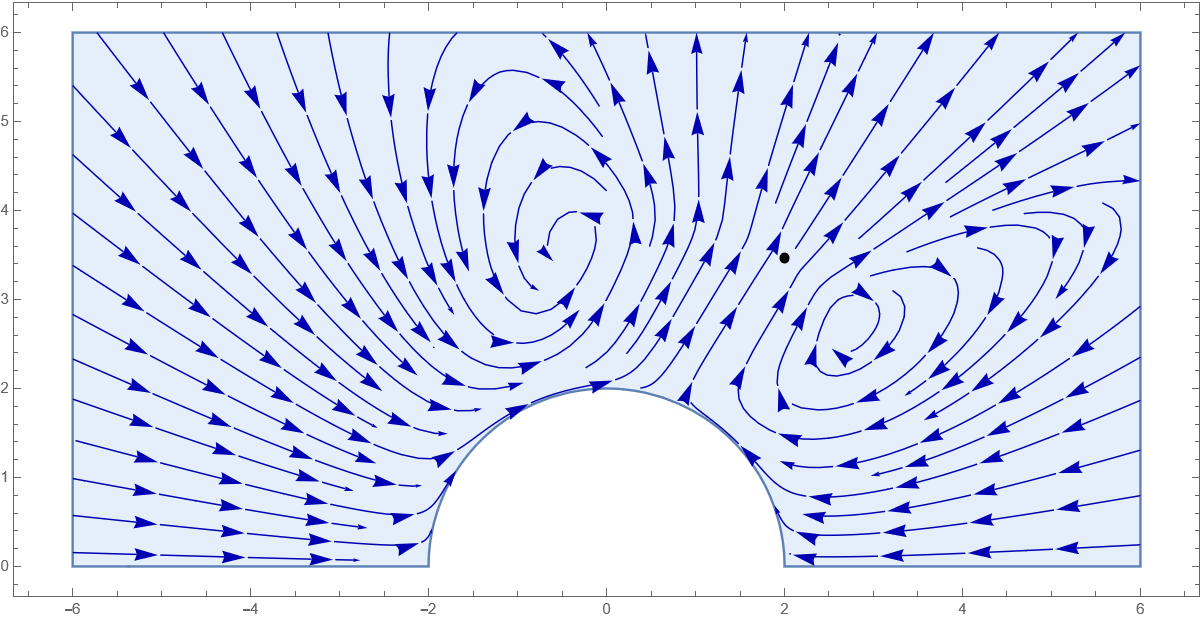}}\quad
\subfloat[Integral curves of $\vec{B}$.]{\includegraphics[width=.4\textwidth,keepaspectratio]{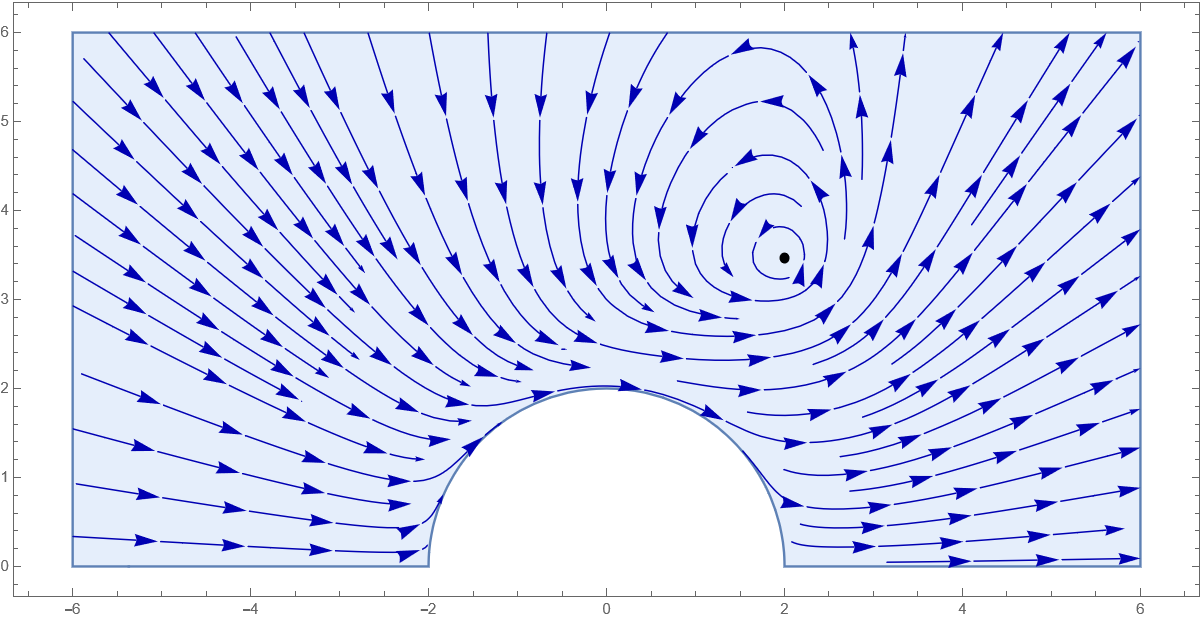}}
\caption{Integral curves of the electric and the magnetic field around the current loop (depicted by black dot) above the black hole as measured by ZAMO in the $(r,\theta)$ plane. The rotational axis is horizontal and the parameters are $r_p=2,r_m=1.999,r_0=4,\theta_0=\pi/3$.}
\label{fig:current}
\end{center}
\end{figure}

The integral lines of the electric and magnetic fields which would have been measured by congruence of ZAMO observers are visualized in Fig. \ref{fig:current}.

\section{Conclusions}
We provided compact and closed form of the electromagnetic Debye superpotential for circular sources on the Kerr background. This superpotential is not unique as the necessary discontinuities can be moved to any line connecting the source to infinity/axis if viewed as a one-valued function after ramification. Therefore, also the distributional sources of the Debye potential are not unique, however these have no physical meaning.

Having this superpotential at hand we discussed the field of the charged ring and the circular current loop, these results has been known only in terms of series expansion so far.

We demonstrated that our results are in agreement with previous results obtained in a form of series.

Field of charged ring and of circular current loop can be considered as elementary building blocks for more complicated and astrophysically interesting axially symmetric stationary configurations --- e.g. (slowly) accreting disks around rotating black holes. These can be obtained from our results by numerical integration.

\begin{acknowledgements}
        D.K. and P.K. acknowledge the support from the Czech Science Foundation, Grant 21-11268S. The calculations were performed using \emph{Wolfram Mathematica}$^{\circledR}$ as well as \emph{MapleSoft Maple}$^{\circledR}$.
        
        D.K. is also deeply indebted to Old\v{r}ich Semer\'a{k} whose comments helped to improve the manuscript.
        
        \textbf{Data Availability Statement:} Data sharing not applicable to this article as no datasets were generated or analyzed during the current study.

\end{acknowledgements}

\appendix
\section{NP and GHP formalism}
\label{app:NPGHP}
In the NP formalism \cite{Newman1962} quantities describing spacetime geometry and field equations are expressed in terms of scalars obtained as their projections onto the null tetrad $(\vec{l},\,\vec{m},\,\vec{\bar{m}}\,,\vec{n})$. The tetrad is determined by demanding that the only nonvanishing scalar products are $\vec{l}^a\vec{n}_a=-\vec{m}^a\vec{\bar{m}}_a=-1$ (in our adopted signature convention $(-,+,+,+)$).
The freedom in its choice is given by the Lorentz group which naturally splits into 4 groups: null rotations around fixed $\vec{l}$, null rotations around fixed $\vec{n}$, boosts in $(\vec{l},\,\vec{n})$ plane, and rotations in $(\vec{m},\,\vec{\bar{m}})$ plane. The metric is reconstructed as $\vec{g}_{ab}=-2\vec{l}_{(a}\vec{n}_{b)}+2\vec{m}_{(a}\vec{\bar{m}}_{b)}$.
And the connection is encoded in 12 complex spin coefficients. 

The directional derivatives associated with the null tetrad are defined as
\begin{equation}
\begin{aligned}
D&\equiv \vec{l}^a\nabla_{a}\,, &
\delta &\equiv \vec{m}^a\nabla_{a}\,, \\
\Delta &\equiv \vec{n}^a\nabla_{a}\,, &
\bar{\delta} &\equiv \vec{\bar{m}}^a\nabla_{a}\,.
\end{aligned}
\end{equation}

The discrete `prime' transformation which interchanges the null basis vectors as
\begin{align}
        \vec{l}&\stackrel{'}{\longleftrightarrow} \vec{n}\,, &
        \vec{m}&\stackrel{'}{\longleftrightarrow} \vec{\bar{m}}\,,
\label{eq:prime}
\end{align}
allows us to reduce the number of Greek letters needed for spin coefficients, and it is common to use $(\kappa,\,\sigma,\,\rho,\,\tau,\,\beta,\,\epsilon)$ and their primed counterparts $(\kappa'=-\nu,\,\sigma'=-\lambda,\,\rho'=-\mu,\,\tau'=-\pi,\,\beta'=-\alpha,\,\epsilon'=-\gamma)$. These spin coefficients are defined as
\begin{equation}
\begin{aligned}
\kappa &= -\vec{m}^aD\vec{l}_a\,, &
\sigma &= -\vec{m}^a\delta\vec{l}_a\,, \\
\rho &= -\vec{m}^a\bar{\delta}\vec{l}_a\,, &
\tau &= -\vec{m}^a\Delta\vec{l}_a\,,
\end{aligned}
\end{equation}
and
\begin{equation}
\begin{aligned}
\beta &= +\frac{1}{2}\left(\vec{n}^a\delta\vec{l}_a-\vec{\bar{m}}^a\delta\vec{m}_a\right),\\
\epsilon &= -\frac{1}{2}\left(\vec{n}^aD\vec{l}_a-\vec{\bar{m}}^aD\vec{m}_a\right).
\end{aligned}
\end{equation}
The primed counterparts are obtained by prime operation which was defined in (\ref{eq:prime}).

In GHP formalism \cite{Geroch1973} the real null directions $\vec{l},\,\vec{n}$ are fixed and the freedom of the tetrad is restricted to boosts in $(\vec{l},\,\vec{n})$ plane and rotations in $(\vec{m},\,\vec{\bar{m}})$ plane which can be written explicitly as\footnote{This transformation naturally follows from the transformation of spin dyad $o^A\rightarrow \lambda o^A,\,\iota^A\rightarrow \lambda^{-1}\iota^A$.}
\begin{equation}
\begin{aligned}
        \vec{l}^a &\ \rightarrow\ \lambda\bar{\lambda}\, \vec{l}^a,\, &
        \vec{n}^a &\ \rightarrow\ \lambda^{-1}\bar{\lambda}^{-1}\,\vec{n}^a\,, \\
        \vec{m}^a &\ \rightarrow\ \lambda\bar{\lambda}^{-1}\,\vec{m}^a,\, &
        \vec{\bar{m}}^a &\ \rightarrow\ \lambda^{-1}\bar{\lambda}\,\vec{\bar{m}}^a\,,
\end{aligned}\label{eq:brt}
\end{equation}
where $\lambda$ is an arbitrary nonvanishing complex function.
This allows us to define GHP scalar of a specific weight $[p,q]$ (corresponding to a spin- and boost-weight $(\frac{1}{2}(p-q),\,\frac{1}{2}(p+q))$) which transforms as 
\begin{equation}
        \phi \longrightarrow \lambda^p\bar{\lambda}^q\, \phi\,,
\end{equation}
under the transformations \re{brt}. The $(\kappa,\,\sigma,\,\rho,\,\tau)$\footnote{Together with their primed counterparts.\label{fn:prime}} are proper GHP scalars meanwhile neither $(\beta,\,\epsilon)^{\ref{fn:prime}}$ nor NP directional derivatives $(D,\,\Delta,\,\delta,\,\bar{\delta})$ transform properly. Incorporating $\beta$ and $\epsilon$ in differential operators leads to GHP derivatives
\begin{equation}
\begin{aligned}
        \thorn \eta &= \left(D-p\epsilon-q\bar{\epsilon}\right)\eta\,, &
        \thorn' \eta &= \left(\Delta+p\epsilon'+q\bar{\epsilon}'\right)\eta\,, \\
        \eth \eta &=\left(\delta-p\beta+q\bar{\beta}'\right)\eta \,, &
        \eth' \eta &=\left(\bar{\delta}+p\beta'-q\bar{\beta}\right)\eta \,,
\end{aligned}
\end{equation}
which acting on scalar of weight $[p,q]$ create a scalar of weight $[p+r,q+s]$ where the appropriate raising/lowering weights $[r,s]$ of the particular derivative are as follows 
\begin{equation}
\begin{aligned}
        \thorn &\rightarrow [+1,+1]\,, &\thorn' &\rightarrow [-1,-1]\,, \\
        \eth &\rightarrow [+1,-1]\,,  &\eth' &\rightarrow [-1,+1]\,.
\end{aligned}
\end{equation}
Therefore $\eth$ and $\eth'$ are spin raising and lowering operators, meanwhile $\thorn$ and $\thorn'$ are boost raisining and lowering operators.

The prime operation takes a scalar of weight $[p,q]$ into a scalar of weight $[-p,-q]$, and complex conjugation into a scalar of weight $[q,p]$.

The GHP formalism allows for a simple consistency test of equations: only a scalars of the same GHP weights can be compared. 

The Weyl tensor $\vec{C}_{abcd}$ is encoded in 5 complex scalars $\psi_j$, $j\in(0,1,2,3,4)$. In spacetimes of algebraic type D in the aligned tetrad only $\psi_2$ is nonzero
\begin{equation}
\psi_2=\vec{C}_{abcd}\vec{l}^a\vec{m}^b\vec{\bar{m}}^c\vec{n}^d \,,
\end{equation}
whereas for Maxwell tensor $\vec{F}_{\!ab}$ we have 3 complex scalars
\begin{equation}
\begin{aligned}
\phi_0 &= \vec{F}_{\!ab}\vec{l}^a\vec{m}^b\,, \\
\phi_1 &= \frac{1}{2}\left(\vec{F}_{\!ab}\vec{l}^a\vec{n}^b-\vec{m}^a\vec{\bar{m}}^b\right), \\
\phi_2 &= \vec{F}_{\!ab}\vec{\bar{m}}^a\vec{n}^b\,.
\end{aligned}
\end{equation}

\section{NP quantities of Kerr black hole}
\label{app:Kerr}
The nonzero NP spin coefficients corresponding to the tetrad (\ref{eq:NPtetrad}) are 
\begin{equation}
\begin{alignedat}{2}
\pi &= \frac{-i}{\sqrt{2}}\,\frac{\ka \sin\theta}{\Krho^2}\,,\qquad &
\mu &= \frac{1}{\sqrt{2}}\,\frac{\Delta}{\Sigma\Krho}\,, \\
\tau &= \frac{i}{\sqrt{2}}\,\frac{\ka \sin\theta}{\Sigma}\,, &
\rho &= \frac{1}{\sqrt{2}}\,\frac{1}{\Krho}\,,\\
\gamma &=\mu -\frac{1}{\sqrt{2}}\,\frac{r-M}{\Sigma}\,,\qquad &
\beta &= \frac{-1}{2\sqrt{2}}\,\frac{\cot \theta}{\Krhocc}\,, \\
\alpha &= \pi-\bar{\beta}\,, & 
\end{alignedat}\label{eq:spinc}
\end{equation}
and the only nonzero Weyl scalar reads 
\begin{equation}
\psi_2 = \frac{M}{\Krho^3} \,.
\label{eq:psi2}\end{equation}

\section{ZAMO congruence}
\label{app:ZAMO}
The physical interpretation of the electromagnetic field is done by an observer who makes a local measurements. Physical measurements in GR are done by projections of the field onto an orthonormal tetrad. One of the most useful congruence of observers around the Kerr black hole are the ZAMO observers whose 4-velocity is defined by $\vec{u}_a\propto (\vec{d} t)_a$, the congruence is thus non-twisting and, as its name suggests, angular momentum of every particular observer vanishes, i.e. $L\equiv\vec{\eta}\cdot\vec{u}=0$. The tetrad ($\vec{u}\equiv\vec{e}_{(t)}$) is given by
\begin{equation}
\begin{alignedat}{2}
\vec{e}_{(t)} &= \frac{1}{N} \left( \vec{\p_t}+\omega\,\vec{\p_\phi} \right),\quad &
\vec{e}_{(r)} &= \sqrt{\frac{\Delta}{\Sigma}}\,\vec{\p_r}\,, \\
\vec{e}_{(\theta)} &= \frac{1}{\sqrt{\Sigma}}\,\vec{\p_\theta}\,,&
\vec{e}_{(\phi)} &= \frac{1}{\sin\theta}\,\sqrt{\frac{\Sigma}{\Upsilon}}\,\vec{\p_\phi}\,,
\end{alignedat}
\label{eq:ZAMO}
\end{equation}
where
\begin{align}
N &= \sqrt{ \frac{\left( \vec{\eta}\cdot\vec{\xi} \right)^2}{ \vec{\eta}\cdot\vec{\eta} }-\vec{\xi}\cdot\vec{\xi}}\,, \\
\omega &=-\frac{\vec{\xi}\cdot\vec{\eta}}{\vec{\eta}\cdot\vec{\eta}}\,,\\
\Upsilon &=\Delta\Sigma+r(r_p+r_m)\left(r^2+r_p r_m\right)\,.
\end{align}
The scalar product of two vectors $\vec{u}$ and $\vec{v}$ is denoted as $\vec{u}\cdot\vec{v}=\vec{u}^a\vec{g}_{ab}\vec{v}^b$. The Killing vectors of the Kerr metric are $\vec{\xi}=\vec{\p_t}$ and $\vec{\eta}=\vec{\p_\phi}$.
The projections 
\begin{equation}
\mathcal{E}_{(k)} = \vec{e}_{(t)}\cdot\vec{F}^*\cdot\vec{e}_{(k)}\,,\qquad\text{for}\qquad k\in (r,\,\theta,\,\phi)\,,
\end{equation}
written in compact form for $\vec{\mathcal{E}}=\vec{E}-i\vec{B}$ are
\begin{align}
\mathcal{E}_{(r)} &= \frac{\frac{-i \ka\sin\theta\Delta}{\vrho}\,\phi_0-2\left(r^2+\ka^2\right)\phi_1+i\ka\sin\theta\,\vrho\,\phi_2}{\sqrt{\Upsilon}} \,, \nonumber\\
\mathcal{E}_{(\theta)} &= \frac{r^2+\ka^2}{\sqrt{\Delta\Upsilon}}
    \left( \frac{\Delta}{\vrho}\,\phi_0-\frac{2i\ka\sin\theta\Delta}{r^2+\ka^2}\,\phi_1-\bar{\rho}\,\phi_2\right),\nonumber\\
\mathcal{E}_{(\phi)} &= -\frac{i\sqrt{\Delta}}{\vrho}\,\phi_0-\frac{i\vrho}{\sqrt{\Delta}}\,\phi_2\,.
\label{eq:projE}
\end{align}

\section{Elliptic integrals}
\label{app:ellitptic}
We use the same definition of complete elliptic integrals as the one implemented in \emph{Wolfram Mathematica}$^{\circledR}$, i.e.
\begin{align}
E(m) &= \int_0^{\pi/2}\sqrt{1-m\sin^2\theta}\,\d\theta\,, \\
K(m) &= \int_0^{\pi/2}\frac{1}{\sqrt{1-m\sin^2\theta}}\,\d\theta\,, \\
\Pi(n|m) &= \int_0^{\pi/2}\frac{1}{\left(1-n\sin^2\theta\right)\sqrt{1-m\sin^2\theta}}\,\d\theta\,.
\end{align}

\section{Maxwell Equations in axisymmetric stationary case}
\label{app:ME}
Maxwell Equations in axisymmetric stationary case can be, using the the following rescaled NP quantities
\begin{align}
    \phi_0 &= \frac{\sqrt{2}\vrho}{\sin\theta} \, \tilde{\phi}_0 \,, &
    \phi_1 &= \frac{\sqrt{2}}{\vrho^2} \, \tilde{\phi}_1\,,&
    \phi_2 &= \frac{\sqrt{2}\Delta}{\vrho\sin\theta} \, \tilde{\phi}_2 \,,
\end{align}
rewritten as
\begin{align}
    \frac{ \frac{\p}{\p\theta}\,\tilde{\phi}_0 }{\sin\theta}  
    - \frac{ \frac{\p}{\p r}\, \tilde{\phi}_1}{\vrho^2} &=-J_l\,, &
    \frac{ \frac{\p}{\p r}\,\left(\Delta\tilde{\phi}_0\right)}{\bar{\vrho}\sin\theta}  
    + \frac{\frac{\p}{\p \theta}\, \tilde{\phi}_1}{\vrho\Sigma}   &=J_m\,,\\
    \frac{ \frac{\p}{\p\theta}\,\tilde{\phi}_2 }{\Sigma\sin\theta}  
    + \frac{ \Delta\frac{\p}{\p r}\, \tilde{\phi}_1 }{\vrho^2\Sigma} &=J_n\,, &
    \frac{ \frac{\p}{\p r}\,\tilde{\phi}_2 }{\vrho\sin\theta}  
     - \frac{\frac{\p}{\p \theta}\, \tilde{\phi}_1}{\vrho^3}   &=J_{\bar{m}}\,.
\end{align}

\section{Different discontinuities location}
\label{app:discont}

We may express
\begin{equation}
\begin{alignedat}{2}
r &=\frac{1}{2}\left(r_p+r_m+\sqrt{(z-\beta)^2+\rho^2}+\sqrt{(z+\beta)^2+\rho^2}\right)\,, \\
r_0 &=\frac{1}{2}\left(r_p+r_m+\sqrt{(\zz-\beta)^2+\rz^2}+\sqrt{(\zz+\beta)^2+\rz^2}\right)
\end{alignedat}
\end{equation}
and then define potentials
\begin{align}
\Xi_\textbf{0} &= \Xi_\textbf{r}
-\frac{4i\pi \mathcal{R}_+}{\beta\sin\theta_0}\,\Xi_\textbf{n}\Theta_\textbf{n} \nonumber\\
&\quad -\left(\frac{4i\pi(r_p+r_m)}{\sin\theta_0}\,\Xi_\textbf{i} + \frac{4i\pi\mathcal{R}_+}{\beta\sin\theta_0}\,\Xi_\textbf{n}\right)\Theta_\textbf{s}\,, \\
\Xi_\textbf{1} &= \Xi_\textbf{r}
-\frac{2i\pi\mathcal{R}_+}{\beta\sin\theta_0}\,\Xi_\textbf{n}\Theta_\textbf{n}\left(1+\Theta(r_0-r)\right) \nonumber\\
&\quad -\left(\frac{4i\pi(r_p+r_m)}{\sin\theta_0}\,\Xi_\textbf{i} + \frac{2i\pi\mathcal{R}_+}{\beta\sin\theta_0}\,\Xi_\textbf{n}\right)\Theta_\textbf{s} \,,\\
\Xi_\textbf{2} &= \Xi_\textbf{r} 
-\frac{2i\pi\mathcal{R}_+}{\beta\sin\theta_0}\,\Xi_\textbf{n}\Theta_\textbf{n}\left(1+\Theta(r_0-r)\right) \nonumber\\
&\quad -\left(\frac{4i\pi(r_p+r_m)}{\sin\theta_0}\,\Xi_\textbf{i} + \frac{2i\pi\mathcal{R}_+}{\beta\sin\theta_0}\,\Xi_\textbf{n}\right)\Theta_\textbf{s} \Theta(r-r_0)\nonumber\\
&\quad +\frac{4i\pi\mathcal{R}_-}{\beta\sin\theta_0} \Theta(r_0-r) \Theta_\textbf{s} \Xi_\textbf{s} \,,
\end{align}
where
\begin{equation}
\begin{alignedat}{2}
\mathcal{R}_+ &= (\beta+i\ka)\sqrt{(\zz-\beta)^2+\rz^2}\,, \\
\mathcal{R}_- &= (\beta-i\ka)\sqrt{(\zz+\beta)^2+\rz^2}\,. 
\end{alignedat}
\end{equation}
All of these potential represent the same field. They differ just by the position of discontinuities --- visualisations can be found in Fig \ref{fig:jumps2}. And, moreover, there is still a great freedom since one can add terms
\begin{align}
a_\textbf{n} \Xi_\textbf{n}+a_\textbf{i} \Xi_\textbf{i}+a_\textbf{s} \Xi_\textbf{s}\,, &&
a_\textbf{n}+a_\textbf{i}+a_\textbf{s} &=0\,,
\end{align}
i.e. Debye potential of field representing monopole with vanishing charge.

\section{Riemann surface}
\label{app:Riem}
\begin{figure}
\begin{center}
\includegraphics[keepaspectratio,width=.3\textwidth]{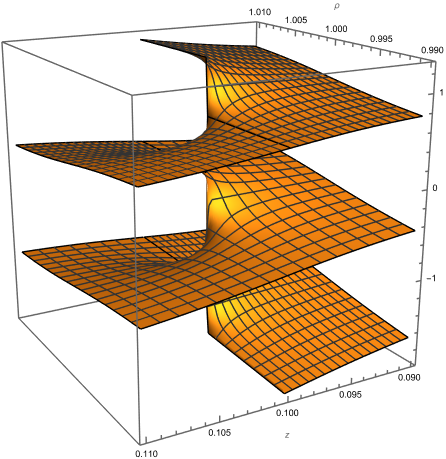}
\caption{Riemann surface of function $u+jv$ for $\zz=0.1,\,\rz=1$, the values of $\mathrm{Re}(u+jv)$ are on vertical axis.}
\label{fig:rs}
\end{center}
\end{figure}

Let us visualize the Riemann surface of function (the third term in Eq. (\ref{eq:f}))
\begin{equation}
u=-\frac{4(z+\zz)\rz}{d(\zz,\rz)\rho^2}\,\Pi\left(\frac{h(\zz,-\rz)}{h(\zz,\rz)},\mu'(\rz)\right),
\end{equation}
which can be analytically continued across $\vec{\gamma}_\textbf{i}$ by adding
\begin{equation}
v=-\frac{2\pi(z+\zz)}{\rho^2}\,
\end{equation}
and in general we have an infinite number of sheets
\begin{equation}
u+jv\,\qquad j\in \mathbb{Z}\,.
\end{equation}
The appropriate Riemann surface is in Fig. \ref{fig:rs} and clearly shows the branch point at $(\zz,\rz)$. To choose a branch cut is to join a branch point with axis or infinity and choose a particular sheet.

It is easy to write down analytical continuation of the Debye potential $\Xi$ as given in (\ref{eq:XiSmooth}) itselt, but it is ``complicated'' to make a representative visualisation of its Riemann surfaces since the discontinuities are relatively small. We just have to add
\begin{equation}
\begin{aligned}
&\frac{\pi}{\rho^2\sin\theta_0}\Bigl[ 2(z+\zz)\\
&\quad +\sqrt{\left((z-\beta)^2+\rho^2\right)\left((\zz-\beta)^2+\rz^2\right)}/\beta \\
&\qquad -\sqrt{\left((z+\beta)^2+\rho^2\right)\left((\zz+\beta)^2+\rz^2\right)}/\beta\Bigr]\,,
\end{aligned}
\end{equation}
to the Debye potential $\Xi$.


%


\begin{thebibliography}{28}%
\makeatletter
\providecommand \@ifxundefined [1]{%
 \@ifx{#1\undefined}
}%
\providecommand \@ifnum [1]{%
 \ifnum #1\expandafter \@firstoftwo
 \else \expandafter \@secondoftwo
 \fi
}%
\providecommand \@ifx [1]{%
 \ifx #1\expandafter \@firstoftwo
 \else \expandafter \@secondoftwo
 \fi
}%
\providecommand \natexlab [1]{#1}%
\providecommand \enquote  [1]{``#1''}%
\providecommand \bibnamefont  [1]{#1}%
\providecommand \bibfnamefont [1]{#1}%
\providecommand \citenamefont [1]{#1}%
\providecommand \href@noop [0]{\@secondoftwo}%
\providecommand \href [0]{\begingroup \@sanitize@url \@href}%
\providecommand \@href[1]{\@@startlink{#1}\@@href}%
\providecommand \@@href[1]{\endgroup#1\@@endlink}%
\providecommand \@sanitize@url [0]{\catcode `\\12\catcode `\$12\catcode
  `\&12\catcode `\#12\catcode `\^12\catcode `\_12\catcode `\%12\relax}%
\providecommand \@@startlink[1]{}%
\providecommand \@@endlink[0]{}%
\providecommand \url  [0]{\begingroup\@sanitize@url \@url }%
\providecommand \@url [1]{\endgroup\@href {#1}{\urlprefix }}%
\providecommand \urlprefix  [0]{URL }%
\providecommand \Eprint [0]{\href }%
\providecommand \doibase [0]{https://doi.org/}%
\providecommand \selectlanguage [0]{\@gobble}%
\providecommand \bibinfo  [0]{\@secondoftwo}%
\providecommand \bibfield  [0]{\@secondoftwo}%
\providecommand \translation [1]{[#1]}%
\providecommand \BibitemOpen [0]{}%
\providecommand \bibitemStop [0]{}%
\providecommand \bibitemNoStop [0]{.\EOS\space}%
\providecommand \EOS [0]{\spacefactor3000\relax}%
\providecommand \BibitemShut  [1]{\csname bibitem#1\endcsname}%
\let\auto@bib@innerbib\@empty
\bibitem [{\citenamefont {Kerr}(1963)}]{Kerr1963}%
  \BibitemOpen
  \bibfield  {author} {\bibinfo {author} {\bibfnamefont {R.~P.}\ \bibnamefont
  {Kerr}},\ }\bibfield  {title} {\bibinfo {title} {{Gravitational Field of a
  Spinning Mass as an Example of Algebraically Special Metrics}},\ }\href
  {https://doi.org/10.1103/PhysRevLett.11.237} {\bibfield  {journal} {\bibinfo
  {journal} {Phys. Rev. Lett.}\ }\textbf {\bibinfo {volume} {11}},\ \bibinfo
  {pages} {237} (\bibinfo {year} {1963})}\BibitemShut {NoStop}%
\bibitem [{\citenamefont {Punsly}(2009)}]{Punsly2009}%
  \BibitemOpen
  \bibfield  {author} {\bibinfo {author} {\bibfnamefont {B.}~\bibnamefont
  {Punsly}},\ }\href {https://doi.org/10.1007/978-3-540-76957-6} {\emph
  {\bibinfo {title} {{Black Hole Gravitohydromagnetics}}}},\ \bibinfo {series}
  {Astrophysics and Space Science Library}, Vol.\ \bibinfo {volume} {355}\
  (\bibinfo  {publisher} {Springer Berlin Heidelberg},\ \bibinfo {address}
  {Berlin, Heidelberg},\ \bibinfo {year} {2009})\BibitemShut {NoStop}%
\bibitem [{\citenamefont {Newman}\ and\ \citenamefont
  {Penrose}(1962)}]{Newman1962}%
  \BibitemOpen
  \bibfield  {author} {\bibinfo {author} {\bibfnamefont {E.}~\bibnamefont
  {Newman}}\ and\ \bibinfo {author} {\bibfnamefont {R.}~\bibnamefont
  {Penrose}},\ }\bibfield  {title} {\bibinfo {title} {{An Approach to
  Gravitational Radiation by a Method of Spin Coefficients}},\ }\href
  {https://doi.org/10.1063/1.1724257} {\bibfield  {journal} {\bibinfo
  {journal} {J. Math. Phys.}\ }\textbf {\bibinfo {volume} {3}},\ \bibinfo
  {pages} {566} (\bibinfo {year} {1962})}\BibitemShut {NoStop}%
\bibitem [{\citenamefont {Geroch}\ \emph {et~al.}(1973)\citenamefont {Geroch},
  \citenamefont {Held},\ and\ \citenamefont {Penrose}}]{Geroch1973}%
  \BibitemOpen
  \bibfield  {author} {\bibinfo {author} {\bibfnamefont {R.}~\bibnamefont
  {Geroch}}, \bibinfo {author} {\bibfnamefont {A.}~\bibnamefont {Held}},\ and\
  \bibinfo {author} {\bibfnamefont {R.}~\bibnamefont {Penrose}},\ }\bibfield
  {title} {\bibinfo {title} {{A space-time calculus based on pairs of null
  directions}},\ }\href {https://doi.org/10.1063/1.1666410} {\bibfield
  {journal} {\bibinfo  {journal} {J. Math. Phys.}\ }\textbf {\bibinfo {volume}
  {14}},\ \bibinfo {pages} {874} (\bibinfo {year} {1973})},\ \Eprint
  {https://arxiv.org/abs/1011.1669v3} {arXiv:1011.1669v3} \BibitemShut
  {NoStop}%
\bibitem [{\citenamefont {Teukolsky}(1972)}]{Teukolsky1972}%
  \BibitemOpen
  \bibfield  {author} {\bibinfo {author} {\bibfnamefont {S.~A.}\ \bibnamefont
  {Teukolsky}},\ }\bibfield  {title} {\bibinfo {title} {{Rotating black holes:
  Separable wave equations for gravitational and electromagnetic
  perturbations}},\ }\href {https://doi.org/10.1103/PhysRevLett.29.1114}
  {\bibfield  {journal} {\bibinfo  {journal} {Phys. Rev. Lett.}\ }\textbf
  {\bibinfo {volume} {29}},\ \bibinfo {pages} {1114} (\bibinfo {year}
  {1972})}\BibitemShut {NoStop}%
\bibitem [{\citenamefont {Teukolsky}(1973)}]{Teukolsky1973}%
  \BibitemOpen
  \bibfield  {author} {\bibinfo {author} {\bibfnamefont {S.~A.}\ \bibnamefont
  {Teukolsky}},\ }\bibfield  {title} {\bibinfo {title} {{Perturbations of a
  Rotating Black Hole. I. Fundamental Equations for Gravitational,
  Electromagnetic, and Neutrino-Field Perturbations}},\ }\href
  {https://doi.org/10.1086/152444} {\bibfield  {journal} {\bibinfo  {journal}
  {Astrophys. J.}\ }\textbf {\bibinfo {volume} {185}},\ \bibinfo {pages} {635}
  (\bibinfo {year} {1973})}\BibitemShut {NoStop}%
\bibitem [{\citenamefont {Fackerell}\ and\ \citenamefont
  {Ipser}(1972)}]{Fackerell1972}%
  \BibitemOpen
  \bibfield  {author} {\bibinfo {author} {\bibfnamefont {E.~D.}\ \bibnamefont
  {Fackerell}}\ and\ \bibinfo {author} {\bibfnamefont {J.~R.}\ \bibnamefont
  {Ipser}},\ }\bibfield  {title} {\bibinfo {title} {{Weak Electromagnetic
  Fields Around a Rotating Black Hole}},\ }\href
  {https://doi.org/10.1103/PhysRevD.5.2455} {\bibfield  {journal} {\bibinfo
  {journal} {Phys. Rev. D}\ }\textbf {\bibinfo {volume} {5}},\ \bibinfo {pages}
  {2455} (\bibinfo {year} {1972})}\BibitemShut {NoStop}%
\bibitem [{\citenamefont {Jezierski}\ and\ \citenamefont
  {Smo{\l}ka}(2016)}]{Jezierski2016}%
  \BibitemOpen
  \bibfield  {author} {\bibinfo {author} {\bibfnamefont {J.}~\bibnamefont
  {Jezierski}}\ and\ \bibinfo {author} {\bibfnamefont {T.}~\bibnamefont
  {Smo{\l}ka}},\ }\bibfield  {title} {\bibinfo {title} {{A geometric
  description of Maxwell field in a Kerr spacetime}},\ }\href
  {https://doi.org/10.1088/0264-9381/33/12/125035} {\bibfield  {journal}
  {\bibinfo  {journal} {Class. Quantum Gravity}\ }\textbf {\bibinfo {volume}
  {33}},\ \bibinfo {pages} {125035} (\bibinfo {year} {2016})},\ \Eprint
  {https://arxiv.org/abs/1502.00599} {arXiv:1502.00599} \BibitemShut {NoStop}%
\bibitem [{\citenamefont {Starobinski}\ and\ \citenamefont
  {Churilov}(1973)}]{Starobinski1973}%
  \BibitemOpen
  \bibfield  {author} {\bibinfo {author} {\bibfnamefont {A.}~\bibnamefont
  {Starobinski}}\ and\ \bibinfo {author} {\bibfnamefont {S.}~\bibnamefont
  {Churilov}},\ }\bibfield  {title} {\bibinfo {title} {{Amplification of
  electromagnetic and gravitational waves scattered by a rotating "black
  hole"}},\ }\href@noop {} {\bibfield  {journal} {\bibinfo  {journal} {Sov. J.
  Exp. Theor. Phys.}\ }\textbf {\bibinfo {volume} {65}},\ \bibinfo {pages} {3}
  (\bibinfo {year} {1973})}\BibitemShut {NoStop}%
\bibitem [{\citenamefont {Press}\ and\ \citenamefont
  {Teukolsky}(1973)}]{Press1973}%
  \BibitemOpen
  \bibfield  {author} {\bibinfo {author} {\bibfnamefont {W.~H.}\ \bibnamefont
  {Press}}\ and\ \bibinfo {author} {\bibfnamefont {S.~A.}\ \bibnamefont
  {Teukolsky}},\ }\bibfield  {title} {\bibinfo {title} {{Perturbations of a
  Rotating Black Hole. II. Dynamical Stability of the Kerr Metric}},\ }\href
  {https://doi.org/10.1086/152445} {\bibfield  {journal} {\bibinfo  {journal}
  {Astrophys. J.}\ }\textbf {\bibinfo {volume} {185}},\ \bibinfo {pages} {649}
  (\bibinfo {year} {1973})}\BibitemShut {NoStop}%
\bibitem [{\citenamefont {Teukolsky}\ and\ \citenamefont
  {Press}(1974)}]{Teukolsky1974}%
  \BibitemOpen
  \bibfield  {author} {\bibinfo {author} {\bibfnamefont {S.~A.}\ \bibnamefont
  {Teukolsky}}\ and\ \bibinfo {author} {\bibfnamefont {W.~H.}\ \bibnamefont
  {Press}},\ }\bibfield  {title} {\bibinfo {title} {{Perturbations of a
  rotating black hole. III - Interaction of the hole with gravitational and
  electromagnetic radiation}},\ }\href {https://doi.org/10.1086/153180}
  {\bibfield  {journal} {\bibinfo  {journal} {Astrophys. J.}\ }\textbf
  {\bibinfo {volume} {193}},\ \bibinfo {pages} {443} (\bibinfo {year}
  {1974})}\BibitemShut {NoStop}%
\bibitem [{\citenamefont {Petterson}(1974)}]{Petterson1974}%
  \BibitemOpen
  \bibfield  {author} {\bibinfo {author} {\bibfnamefont {J.~A.}\ \bibnamefont
  {Petterson}},\ }\bibfield  {title} {\bibinfo {title} {{Magnetic field of a
  current loop around a schwarzschild black hole}},\ }\href
  {https://doi.org/10.1103/PhysRevD.10.3166} {\bibfield  {journal} {\bibinfo
  {journal} {Phys. Rev. D}\ }\textbf {\bibinfo {volume} {10}},\ \bibinfo
  {pages} {3166} (\bibinfo {year} {1974})}\BibitemShut {NoStop}%
\bibitem [{\citenamefont {Chitre}\ and\ \citenamefont
  {Vishveshwara}(1975)}]{Chitre1975}%
  \BibitemOpen
  \bibfield  {author} {\bibinfo {author} {\bibfnamefont {D.~M.}\ \bibnamefont
  {Chitre}}\ and\ \bibinfo {author} {\bibfnamefont {C.~V.}\ \bibnamefont
  {Vishveshwara}},\ }\bibfield  {title} {\bibinfo {title} {{Electromagnetic
  field of a current loop around a Kerr black hole}},\ }\href
  {https://doi.org/10.1103/PhysRevD.12.1538} {\bibfield  {journal} {\bibinfo
  {journal} {Phys. Rev. D}\ }\textbf {\bibinfo {volume} {12}},\ \bibinfo
  {pages} {1538} (\bibinfo {year} {1975})}\BibitemShut {NoStop}%
\bibitem [{\citenamefont {Petterson}(1975)}]{Petterson1975}%
  \BibitemOpen
  \bibfield  {author} {\bibinfo {author} {\bibfnamefont {J.~A.}\ \bibnamefont
  {Petterson}},\ }\bibfield  {title} {\bibinfo {title} {{Stationary
  axisymmetric electromagnetic fields around a rotating black hole}},\ }\href
  {https://doi.org/10.1103/PhysRevD.12.2218} {\bibfield  {journal} {\bibinfo
  {journal} {Phys. Rev. D}\ }\textbf {\bibinfo {volume} {12}},\ \bibinfo
  {pages} {2218} (\bibinfo {year} {1975})}\BibitemShut {NoStop}%
\bibitem [{\citenamefont {Linet}(1976)}]{Linet1976}%
  \BibitemOpen
  \bibfield  {author} {\bibinfo {author} {\bibfnamefont {B.}~\bibnamefont
  {Linet}},\ }\bibfield  {title} {\bibinfo {title} {{Electrostatics and
  magnetostatics in the Schwarzschild metric}},\ }\href
  {https://doi.org/10.1088/0305-4470/9/7/010} {\bibfield  {journal} {\bibinfo
  {journal} {J. Phys. A. Math. Gen.}\ }\textbf {\bibinfo {volume} {9}},\
  \bibinfo {pages} {1081} (\bibinfo {year} {1976})}\BibitemShut {NoStop}%
\bibitem [{\citenamefont {Bi{\v{c}}{\'{a}}k}\ and\ \citenamefont
  {Dvoř{\'{a}}k}(1977)}]{Bicak1977}%
  \BibitemOpen
  \bibfield  {author} {\bibinfo {author} {\bibfnamefont {J.}~\bibnamefont
  {Bi{\v{c}}{\'{a}}k}}\ and\ \bibinfo {author} {\bibfnamefont {L.}~\bibnamefont
  {Dvoř{\'{a}}k}},\ }\bibfield  {title} {\bibinfo {title} {{Stationary
  electromagnetic fields around black holes}},\ }\href
  {https://doi.org/10.1007/BF01587004} {\bibfield  {journal} {\bibinfo
  {journal} {Czechoslov. J. Phys.}\ }\textbf {\bibinfo {volume} {27}},\
  \bibinfo {pages} {127} (\bibinfo {year} {1977})}\BibitemShut {NoStop}%
\bibitem [{\citenamefont {Bi{\v{c}}{\'{a}}k}\ and\ \citenamefont
  {Dvoř{\'{a}}k}(1976)}]{Bicak1976}%
  \BibitemOpen
  \bibfield  {author} {\bibinfo {author} {\bibfnamefont {J.}~\bibnamefont
  {Bi{\v{c}}{\'{a}}k}}\ and\ \bibinfo {author} {\bibfnamefont {L.}~\bibnamefont
  {Dvoř{\'{a}}k}},\ }\bibfield  {title} {\bibinfo {title} {{Stationary
  electromagnetic fields around black holes. II. General solutions and the
  fields of some special sources near a Kerr black hole}},\ }\href
  {https://doi.org/10.1007/BF00766421} {\bibfield  {journal} {\bibinfo
  {journal} {Gen. Relativ. Gravit.}\ }\textbf {\bibinfo {volume} {7}},\
  \bibinfo {pages} {959} (\bibinfo {year} {1976})}\BibitemShut {NoStop}%
\bibitem [{\citenamefont {Cohen}\ and\ \citenamefont
  {Kegeles}(1974{\natexlab{a}})}]{Cohen1974a}%
  \BibitemOpen
  \bibfield  {author} {\bibinfo {author} {\bibfnamefont {J.~M.}\ \bibnamefont
  {Cohen}}\ and\ \bibinfo {author} {\bibfnamefont {L.~S.}\ \bibnamefont
  {Kegeles}},\ }\bibfield  {title} {\bibinfo {title} {{Electromagnetic fields
  in curved spaces: A constructive procedure}},\ }\href
  {https://doi.org/10.1103/PhysRevD.10.1070} {\bibfield  {journal} {\bibinfo
  {journal} {Phys. Rev. D}\ }\textbf {\bibinfo {volume} {10}},\ \bibinfo
  {pages} {1070} (\bibinfo {year} {1974}{\natexlab{a}})}\BibitemShut {NoStop}%
\bibitem [{\citenamefont {Cohen}\ and\ \citenamefont
  {Kegeles}(1974{\natexlab{b}})}]{Cohen1974b}%
  \BibitemOpen
  \bibfield  {author} {\bibinfo {author} {\bibfnamefont {J.}~\bibnamefont
  {Cohen}}\ and\ \bibinfo {author} {\bibfnamefont {L.}~\bibnamefont
  {Kegeles}},\ }\bibfield  {title} {\bibinfo {title} {{Electromagnetic fields
  in general relativity: A constructive procedure}},\ }\href
  {https://doi.org/10.1016/0375-9601(74)90036-X} {\bibfield  {journal}
  {\bibinfo  {journal} {Phys. Lett. A}\ }\textbf {\bibinfo {volume} {47}},\
  \bibinfo {pages} {261} (\bibinfo {year} {1974}{\natexlab{b}})}\BibitemShut
  {NoStop}%
\bibitem [{\citenamefont {Wald}(1978)}]{Wald1978}%
  \BibitemOpen
  \bibfield  {author} {\bibinfo {author} {\bibfnamefont {R.~M.}\ \bibnamefont
  {Wald}},\ }\bibfield  {title} {\bibinfo {title} {{Construction of solutions
  of gravitational, electromagnetic, or other perturbation equations from
  solutions of decoupled equations}},\ }\href
  {https://doi.org/10.1103/PhysRevLett.41.203} {\bibfield  {journal} {\bibinfo
  {journal} {Phys. Rev. Lett.}\ }\textbf {\bibinfo {volume} {41}},\ \bibinfo
  {pages} {203} (\bibinfo {year} {1978})}\BibitemShut {NoStop}%
\bibitem [{\citenamefont {Kegeles}\ and\ \citenamefont
  {Cohen}(1979)}]{Kegeles1979}%
  \BibitemOpen
  \bibfield  {author} {\bibinfo {author} {\bibfnamefont {L.~S.}\ \bibnamefont
  {Kegeles}}\ and\ \bibinfo {author} {\bibfnamefont {J.~M.}\ \bibnamefont
  {Cohen}},\ }\bibfield  {title} {\bibinfo {title} {{Constructive procedure for
  perturbations of spacetimes}},\ }\href
  {https://doi.org/10.1103/PhysRevD.19.1641} {\bibfield  {journal} {\bibinfo
  {journal} {Phys. Rev. D}\ }\textbf {\bibinfo {volume} {19}},\ \bibinfo
  {pages} {1641} (\bibinfo {year} {1979})}\BibitemShut {NoStop}%
\bibitem [{\citenamefont {Andersson}\ \emph {et~al.}(2015)\citenamefont
  {Andersson}, \citenamefont {B{\"{a}}ckdahl},\ and\ \citenamefont
  {Blue}}]{Andersson2015}%
  \BibitemOpen
  \bibfield  {author} {\bibinfo {author} {\bibfnamefont {L.}~\bibnamefont
  {Andersson}}, \bibinfo {author} {\bibfnamefont {T.}~\bibnamefont
  {B{\"{a}}ckdahl}},\ and\ \bibinfo {author} {\bibfnamefont {P.}~\bibnamefont
  {Blue}},\ }\bibfield  {title} {\bibinfo {title} {{Spin geometry and
  conservation laws in the Kerr spacetime}},\ }\href
  {https://doi.org/10.4310/SDG.2015.v20.n1.a8} {\bibfield  {journal} {\bibinfo
  {journal} {Surv. Differ. Geom.}\ }\textbf {\bibinfo {volume} {20}},\ \bibinfo
  {pages} {183} (\bibinfo {year} {2015})},\ \Eprint
  {https://arxiv.org/abs/1504.02069v1} {arXiv:1504.02069v1} \BibitemShut
  {NoStop}%
\bibitem [{\citenamefont {Aksteiner}\ and\ \citenamefont
  {B{\"{a}}ckdahl}(2019)}]{Aksteiner2019}%
  \BibitemOpen
  \bibfield  {author} {\bibinfo {author} {\bibfnamefont {S.}~\bibnamefont
  {Aksteiner}}\ and\ \bibinfo {author} {\bibfnamefont {T.}~\bibnamefont
  {B{\"{a}}ckdahl}},\ }\bibfield  {title} {\bibinfo {title} {{Symmetries of
  linearized gravity from adjoint operators}},\ }\bibfield  {journal} {\bibinfo
   {journal} {J. Math. Phys.}\ }\textbf {\bibinfo {volume} {60}},\ \href
  {https://doi.org/10.1063/1.5092587} {10.1063/1.5092587} (\bibinfo {year}
  {2019}),\ \Eprint {https://arxiv.org/abs/1609.04584} {arXiv:1609.04584}
  \BibitemShut {NoStop}%
\bibitem [{\citenamefont {Linet}(1979)}]{Linet1979}%
  \BibitemOpen
  \bibfield  {author} {\bibinfo {author} {\bibfnamefont {B.}~\bibnamefont
  {Linet}},\ }\bibfield  {title} {\bibinfo {title} {{Stationary axisymmetry
  electromagnetic fields in the Kerr metric}},\ }\href
  {https://doi.org/10.1088/0305-4470/12/6/013} {\bibfield  {journal} {\bibinfo
  {journal} {J. Phys. A. Math. Gen.}\ }\textbf {\bibinfo {volume} {12}},\
  \bibinfo {pages} {839} (\bibinfo {year} {1979})}\BibitemShut {NoStop}%
\bibitem [{\citenamefont {Penrose}\ and\ \citenamefont
  {Rindler}(1984)}]{Penrose1984}%
  \BibitemOpen
  \bibfield  {author} {\bibinfo {author} {\bibfnamefont {R.}~\bibnamefont
  {Penrose}}\ and\ \bibinfo {author} {\bibfnamefont {W.}~\bibnamefont
  {Rindler}},\ }\href {https://doi.org/10.1017/CBO9780511564048} {\emph
  {\bibinfo {title} {{Spinors and Space-Time}}}}\ (\bibinfo  {publisher}
  {Cambridge University Press},\ \bibinfo {year} {1984})\BibitemShut {NoStop}%
\bibitem [{\citenamefont {Linet}(1977)}]{Linet1977}%
  \BibitemOpen
  \bibfield  {author} {\bibinfo {author} {\bibfnamefont {B.}~\bibnamefont
  {Linet}},\ }\bibfield  {title} {\bibinfo {title} {{Stationary axisymmetric
  test fields on a Kerr metric}},\ }\href
  {https://doi.org/10.1016/0375-9601(77)90030-5} {\bibfield  {journal}
  {\bibinfo  {journal} {Phys. Lett. A}\ }\textbf {\bibinfo {volume} {60}},\
  \bibinfo {pages} {395} (\bibinfo {year} {1977})},\ \Eprint
  {https://arxiv.org/abs/1011.1669v3} {arXiv:1011.1669v3} \BibitemShut
  {NoStop}%
\bibitem [{\citenamefont {Wiltshire}\ \emph {et~al.}(2009)\citenamefont
  {Wiltshire}, \citenamefont {Visser},\ and\ \citenamefont
  {Scott}}]{Wiltshire2009}%
  \BibitemOpen
  \bibfield  {author} {\bibinfo {author} {\bibfnamefont {D.~L.}\ \bibnamefont
  {Wiltshire}}, \bibinfo {author} {\bibfnamefont {M.}~\bibnamefont {Visser}},\
  and\ \bibinfo {author} {\bibfnamefont {S.}~\bibnamefont {Scott}},\
  }\href@noop {} {\emph {\bibinfo {title} {{The Kerr spacetime: Rotating black
  holes in general relativity}}}},\ edited by\ \bibinfo {editor} {\bibfnamefont
  {D.~L.}\ \bibnamefont {Wiltshire}}, \bibinfo {editor} {\bibfnamefont
  {M.}~\bibnamefont {Visser}},\ and\ \bibinfo {editor} {\bibfnamefont
  {S.}~\bibnamefont {Scott}}\ (\bibinfo  {publisher} {Cambridge University
  Press},\ \bibinfo {address} {Cambridge},\ \bibinfo {year} {2009})\ p.\
  \bibinfo {pages} {378}\BibitemShut {NoStop}%
\bibitem [{\citenamefont {Teukolsky}(2015)}]{Teukolsky2015}%
  \BibitemOpen
  \bibfield  {author} {\bibinfo {author} {\bibfnamefont {S.~A.}\ \bibnamefont
  {Teukolsky}},\ }\bibfield  {title} {\bibinfo {title} {{The Kerr metric}},\
  }\bibfield  {journal} {\bibinfo  {journal} {Class. Quantum Gravity}\ }\textbf
  {\bibinfo {volume} {32}},\ \href
  {https://doi.org/10.1088/0264-9381/32/12/124006}
  {10.1088/0264-9381/32/12/124006} (\bibinfo {year} {2015}),\ \Eprint
  {https://arxiv.org/abs/1410.2130} {arXiv:1410.2130} \BibitemShut {NoStop}%
\end{thebibliography}
\end{document}